\pdfoutput=1

\documentclass[11pt,letterpaper]{article}
\usepackage[a4paper,
            bindingoffset=0.2in,
            left=1in,
            right=1in,
            top=1in,
            bottom=1in,
            footskip=.25in]{geometry}
\usepackage{tabularx,booktabs}
\usepackage{url}
\usepackage{times}
\usepackage{fullpage}
\usepackage{graphicx}
\usepackage{lipsum}
\newcommand{\junk}[1]{}

\usepackage{color}
\usepackage{subfigure}
\usepackage{amsfonts}
\usepackage[dvipsnames]{xcolor}
\usepackage{tikz}
\usepackage{threeparttable}
\usepackage{caption}
\usepackage{cite}
\usepackage{authblk}

\usepackage{tabularx,booktabs}
\usepackage{url}
\usepackage{fullpage}
\usepackage{lipsum}

\usepackage{hyperref} 
\hypersetup{
        colorlinks,
        breaklinks=true,
        plainpages=false,
        citecolor=blue,
        linkcolor=blue,
        urlcolor=blue,
        bookmarksopen=true,
        bookmarksnumbered=false,
        bookmarksdepth=5
}
\usepackage{amsmath,amssymb,amsfonts}
\usepackage{amsthm}
\usepackage{siunitx}
\usepackage{multirow}
\usepackage{float}
\usepackage{xr}

\usepackage[numbers,sort&compress]{natbib}

\begin{document}
\thispagestyle{empty}
\title{\huge Accurate Prediction of Antibody Function and Structure Using Bio-Inspired Antibody Language Model}
\author[1,2,7]{Hongtai Jing}
\author[2]{Zhengtao Gao}
\author[3]{Sheng Xu}
\author[2,6]{Tao Shen}
\author[2]{Zhangzhi Peng}
\author[2]{Shwai He}
\author[2]{Tao You}
\author[4,5]{Shuang Ye$^*$}
\author[1,2,3,7,8]{Wei Lin$^*$}
\author[2,3]{Siqi Sun\thanks{Corresponding Author. Email: siqisun@fudan.edu.cn, wlin@fudan.edu.cn, shuang\_ye@fudan.edu.cn}}
\affil[1]{\small Institute of Science and Technology for Brain-Inspired Intelligence, Fudan University, Shanghai, 200433, China}
\affil[2]{Research Institute of Intelligent Complex Systems, Fudan University, Shanghai, 200433, China}
\affil[3]{Shanghai AI Laboratory, Shanghai, 200232, China}
\affil[4]{Department of Gynecologic Oncology, Fudan University Shanghai Cancer
Center, Shanghai, 200032, China}
\affil[5]{Department of Oncology, Shanghai Medical College, Fudan University, Shanghai, 200032, China}
\affil[6]{Zelixir Biotech, Shanghai, 201206, China}
\affil[7]{MOE Frontiers Center for Brain Science, Fudan University, Shanghai, 200032, China}
\affil[8]{School of Mathematical Sciences and Shanghai Center for Mathematical Sciences, Fudan University, Shanghai, 200433, China}

\maketitle
\begin{abstract}
In recent decades, antibodies have emerged as indispensable therapeutics for combating diseases, particularly viral infections. However, their development has been hindered by limited structural information and labor-intensive engineering processes.  Fortunately, significant advancements in deep learning methods have facilitated the precise prediction of protein structure and function by leveraging co-evolution information from homologous proteins.  Despite these advances, predicting the conformation of antibodies remains challenging due to their unique evolution and the high flexibility of their antigen-binding regions.  Here, to address this challenge, we present the Bio-inspired Antibody Language Model (BALM). This model is trained on a vast dataset comprising 336 million 40\% non-redundant unlabeled antibody sequences, capturing both unique and conserved properties specific to antibodies. Notably, BALM showcases exceptional performance across four antigen-binding prediction tasks. Moreover, we introduce BALMFold, an end-to-end method derived from BALM, capable of swiftly predicting full atomic antibody structures from individual sequences.
Remarkably, BALMFold outperforms those well-established methods like AlphaFold2, IgFold, ESMFold, and OmegaFold in the antibody benchmark, demonstrating significant potential to advance innovative engineering and streamline therapeutic antibody development by reducing the need for unnecessary trials.
\end{abstract}


\section{Introduction}\label{sec1}
Antibodies are essential immune proteins produced by B cells in response to invasion of foreign substances (known as antigens) such as bacteria, viruses, and other pathogens. These special proteins can accurately recognize antigens with high specificity, and trigger downstream immune reactions to destruct and eliminate the antigens. Therefore, antibodies can be widely used in clinical medicine, e.g., to identify viral infection, or to reactivate T-cell immunity in cancer treatment. High-throughput sequencing technologies targeting antibody repertoire enable fast acquisition of massive antibody sequence data during antibody maturation, which remarkably reshapes our comprehension of humoral immune responses \cite{georgiou2014promise}. However, the development of antibody-based diagnosis and therapeutics remains money- and time-consuming through current wet-lab protocols with insufficient structure information. Using computational methods to predict antibody structures and functions from their sequences could significantly reduce the number of trial-and-error rounds during antibody screening and characterization, and hence largely promote the efficacy of therapeutic antibody development.

The human antibody's structure is Y-shaped, comprising two identical heavy chains and two identical light chains.  The high specificity of antigenic recognition is primarily managed by three complementarity-determining regions (CDRs) located in the fragment of variable ($F_{v}$) at the tips of the antibody. Among these three CDRs, CDR 1 and CDR 2 have relatively low sequence diversity and limited structural conformations. In contrast, the third CDR of the heavy chain (CDR H3) is the most diverse portion and plays a crucial role in recognizing a wide range of antigens \cite{bashford2019analysis,marks2017antibody}. Therefore, computational modeling of CDR H3 structures remains challenging and requires a profound understanding of the available CDR H3 sequences (in the billions) and the limited 3D structural data (only in the tens of thousands).

The advancements in deep learning and natural language processing techniques have led to a tremendous development of protein language models, which hold promise in decoding protein functional properties \cite{Roshan2019, madani2023,esm1b,Meier2021,Prottrans} and predicting protein structure \cite{alphafold2,rosettafold}. These models 
utilize self-supervised learning paradigms on extensive protein sequence datasets, enabling them to extract intrinsic interdependencies and evolutionary traits critical for precise protein structure and function prediction. When it comes to antibodies, the highly conserved structure of an immunoglobulin fold necessitates tailored modeling approaches to distill billions of unlabeled antibody sequences into a universal representation for predicting both structure and function, all without relying on homology search.  Existing methods \cite{antiberta, Antiberty} straightforwardly train transformer-based language models using antibody sequences from the Observed Antibody Space (OAS) \cite{newOAS} dataset without aligning highly conserved residues to particular positions. In addition, around 40\% of the sequences in the OAS dataset are affected by sequencing artifacts \cite{ablang}, potentially impeding the language model's ability to capture contextual information effectively.  Moreover, given that different positions in antibodies exhibit varying evolutionary rates, it becomes crucial for  language models to allocate more attention to learning the distribution of amino acids at those highly diverse positions.

Regarding antibody structure prediction, numerous methods have been developed based on deep learning frameworks \cite{RepertoireBuilder,ABodyBuilder,DeepAB,ablooper,igfold}. AlphaFold2 \cite{alphafold2} and AlphaFold2-Multimer \cite{afm} have notably demonstrated atomic-level accuracy for general protein structures. However, AlphaFold2 heavily relies on multiple sequence alignments (MSAs), which is less helpful for the prediction of highly variable regions, such as CDR loops, in antibodies. Precisely predicting CDR loops' structures from individual antibody sequences remains a significant challenge. Although OmegaFold \cite{omegafold} and ESMFold \cite{esmfold} have shown potential in predicting monomeric protein structures from individual sequences, they may not fully capture the characteristics inherently rooted in antibody structures. IgFold \cite{igfold}, which levearges the pre-trained antibody language model AntiBERTy \cite{Antiberty} and graph networks, offers faster prediction of antibody structures compared to AlphaFold2. However, it was found to be suboptimal in the accurately predicting conformational structures for nanobody CDR H3 loops using the IgFold algorithm \cite{igfold}, even with the inclusion of crystal structure templates as additional information.

In this article, we proposed the Bio-inspired Antibody Language Model (BALM), which effectively captures the inherent rule governing antibody information encoding. This specialized antibody language model holds the potential  to decipher antibody repertoires and offer valuable guidance in therapeutic development efforts.
The main contributions presented in this article are listed as follows.
\begin{itemize}
    \item We conducted through pre-training BALM, making effective use of the distinctive biological properties inherent in antibody sequences.
    \item Biological representations generated by BALM suggest the evolutionary direction of antibodies upon exposure to antigens.
    \item We thoroughly evaluated BALM's performance on various tasks, including antigen-binding prediction, paratope prediction, redundancy prediction during maturation, and binding affinity prediction, which demonstrated state-of-the-art performance in each of these domains.
    \item Leveraging single sequences as inputs, BALMFold outperforms state-of-the-art approaches in terms of accuracy and efficiency when predicting antibody structures.
\end{itemize}

\section{Results}\label{sec2}
\begin{figure}[h!]
	\centering
    \includegraphics[width=\linewidth]{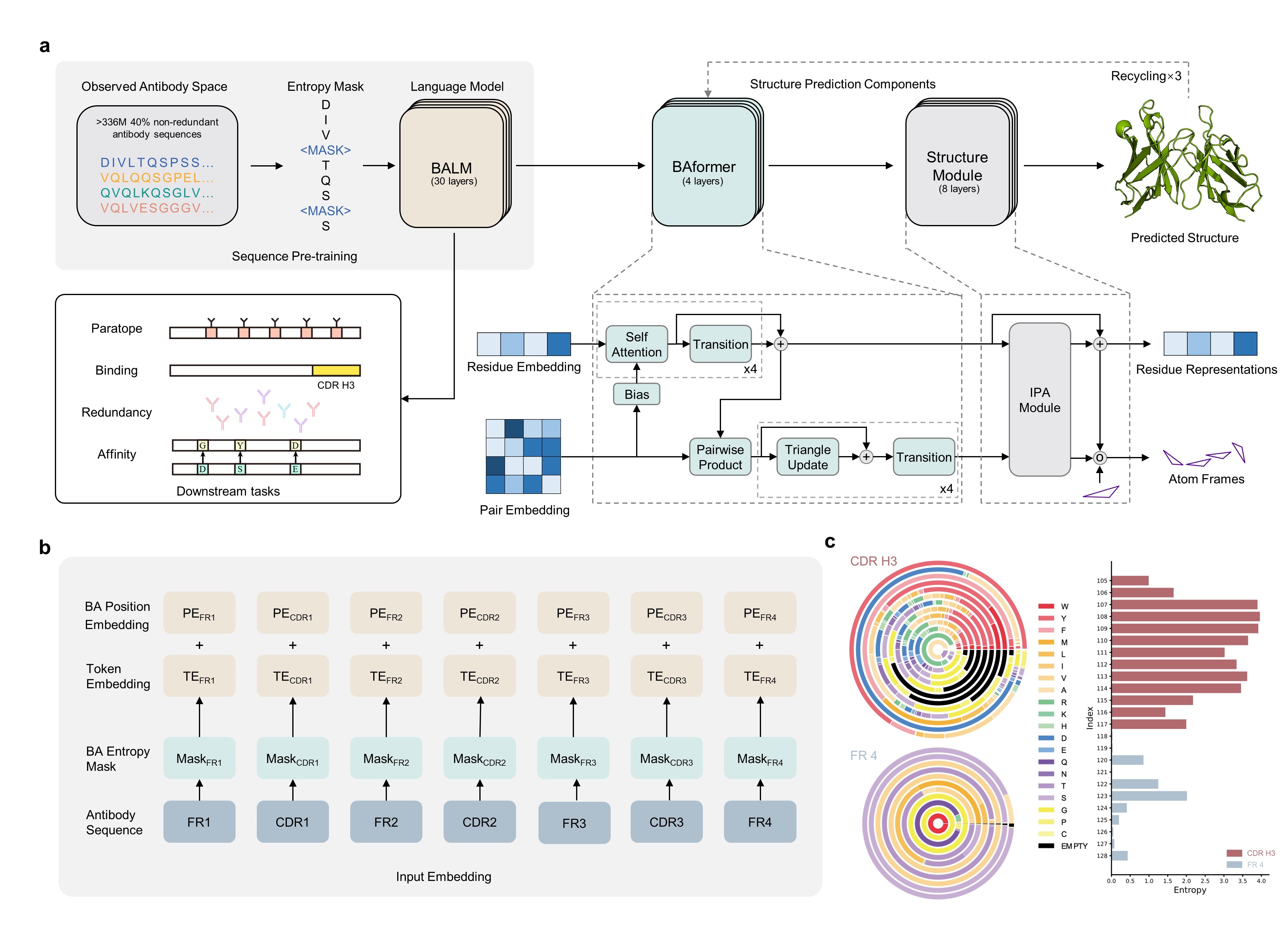}
	\caption{\textbf{Overview of BALM architecture with downstream evaluation.} 
         \textbf{a}, Arrows show the information flow through different blocks. After pre-training on 336M unlabeled antibody sequences, BALM undergoes evaluation for functional properties and structure prediction tasks. The predictions of antigen binding properties encompass tasks related to antigen binding, antibody paratope, antibody affinity, and antibody redundancy. Leveraging the benefits of large-scale sequential antibody data, BALM achieves state-of-the-art performance on each task after fine-tuning. Based on the structural information embeddings of BALM, BALMFold presents an end-to-end approach for full atomic antibody structure prediction. BALMFold comprises 30 transformer layers of BALM, four layers of BAformer, and eight layers of the structure module. Benchmark results demonstrate that BALMFold surpasses alternative methods in both accuracy and speed.
        \textbf{b}, Illustration of the input embedding of BALM. Antibody sequence positions are obtained using the ANARCI tool before being fed into BALM. During pre-training with the masked language modeling objective, residues are masked according to their appearance distribution in each region, enhancing representational learning capability.
        \textbf{c}, The entropy distribution of amino acids in CDR H3 and FR 4 from positions 105 to 128. In the left figure, the variations in the circle's radius correspond to distinct positions. The distribution of amino acids is derived from all heavy chain sequences of paired antibodies obtained from the OAS database.
} 
 \label{fig1}
\end{figure}

\textbf{Leveraging biological features to improve antibody function and structure predictions.}
We trained a bio-inspired, antibody-specific language model, BALM on 336 million unlabeled sequences with 150 million parameters (Fig. \ref{fig1}a). BALM utilizes a transformer-based self-attention mechanism and incorporates a novel antibody positional encoding method (Fig. \ref{fig1}b). Respective residues were subject to masking in accordance with their entropy distribution, and the ensuing task for BALM involved predicting these missing residues (Fig. \ref{fig1}c). All antibody sequences are sourced from the OAS \cite{newOAS} dataset and were clustered using LinClust \cite{LinClust} with a 40\% sequence identity. BALM effectively captures meaningful contextual embeddings of antibody biological features to infer binding function. Leveraging the learned representations of the pre-trained language model, we developed BALMFold as an end-to-end, atomic-level structure prediction algorithm that operates on single sequences. In order to address the challenges of limited homology and the scarcity of structural templates for antibodies, BALMFold employs BALM in conjunction with antibody domain knowledge and a training dataset comprising 2371 paired and 805 single-chain antibody structures with less than $\SI{3}{\angstrom}$ resolution from SAbDab \cite{SAbDab} (before July 1st, 2021) .

To optimally leverage the characteristics inherent in antibodies, we pre-trained BALM with antibody domain knowledge. Unlike common protein sequences, antibody sequences contain distinct regions with varying functions beyond amino acids. The CDR H3 loop is a highly variable region of the antibody sequence that directly interacts with antigens and is a critical component of the antibody sequence, determining specificity and diversity of the antibody repertoire. The frameworks represent the relatively conserved segments of variable domains surrounding the CDRs. Using the IMGT numbering scheme \cite{imgt}, we analyzed the amino acid distribution of heavy chains across 121,838 paired antibody sequences from the OAS database \cite{newOAS}. 

As discernible in the reduced color gradations from CDR H3 to FR 4 in Fig. \ref{fig1}c, our analysis underscores the remarkably variable nature inherent to the CDR H3, as manifested in its colorful segments. In addition, the right side of Fig. \ref{fig1}c demonstrates that the entropy value for CDR H3 (Entropy = 37.1) is considerably, by a factor of seven, superior to that of FR 4 (Entropy = 5.4). The assortment of amino acid combinations plays a crucial role in determining antibody-antigen binding properties. Aligning these positions during the training process can provide valuable biological evolutionary information in the sequences. In contrast to conventional absolute positional encoding, BALM introduces different gaps into distinct antibody sequences based on functional regions. The position ID for each residue is assigned using the antigen receptor numbering and receptor classification (ANARCI) tool \cite{anarci}, following the IMGT \cite{imgt} numbering scheme. To evaluate the effectiveness of antibody functional properties encoded in the antibody language model representation, we conducted experiments on four antigen binding property-related downstream tasks. These tasks encompass antigen binding \cite{Mason2021}, paratope \cite{antiberta}, redundancy prediction \cite{newOAS} during immune response, and binding affinity \cite{SKEMPI,S1131,M1707} between wild type and mutants sequences. By processing aligned antibody sequences, BALM efficiently captures antibody function in biological systems and their interactions with antigens, demonstrating considerable advantages in obtaining antibody-specific representation compared to protein language models \cite{esmfold,Antiberty}.

The language model addresses the challenge of limited training samples in antibody 3D structure prediction.
Multiple sequence alignments (MSAs) offer valuable evolutionary perspectives on the conservation of functionally crucial residues, thereby enhancing the understanding of protein function and structure. For rapidly evolving proteins, MSA-based and template-based approaches have been proved to be unsuitable and time-consuming for antibody domains. The lack of evolutionary information from sequences may impair the performance of evolutionary methods such as AlphaFold2 \cite{alphafold2} and RoseTTAFold \cite{rosettafold}.
To capture the inherent evolutionary information within antibody sequences, we leveraged distinct biological features across various antibody domains using single sequences as input, rather than relying on explicit homologous sequences.
Integrating the potent language model BALM, BALMFold comprises two additional modules: BAformer and the structure module. BAformer updates single and pair representations based on BALM's meaningful representations, while the invariant point attention (IPA) operates in 3D space to generate relative rotations and translations. The structure module produces 3D full atom coordinate structure predictions.

During the training process, we minimized the loss function of BALMFold with respect to four aspects, including frame aligned point error (FAPE), distogram loss, confidence loss, and structure violation loss (additional details in \emph{Methods}). We compared BALMFold with several recent methods \cite{afm, alphafold2, igfold, omegafold, esmfold} in antibody structure prediction and conducted an comparable non-redundant benchmark involving 197 paired antibodies and 71 nanobodies, in alignment with IgFold \cite{igfold}. Without the MSA-based, computationally intensive module, BALMFold generated predicted paired antibody structures within 5.1s and single-chain antibody structures within 2.9s on average during the evaluation of the benchmark.

\begin{figure}[h!]
	\centering
\includegraphics[width=\linewidth]{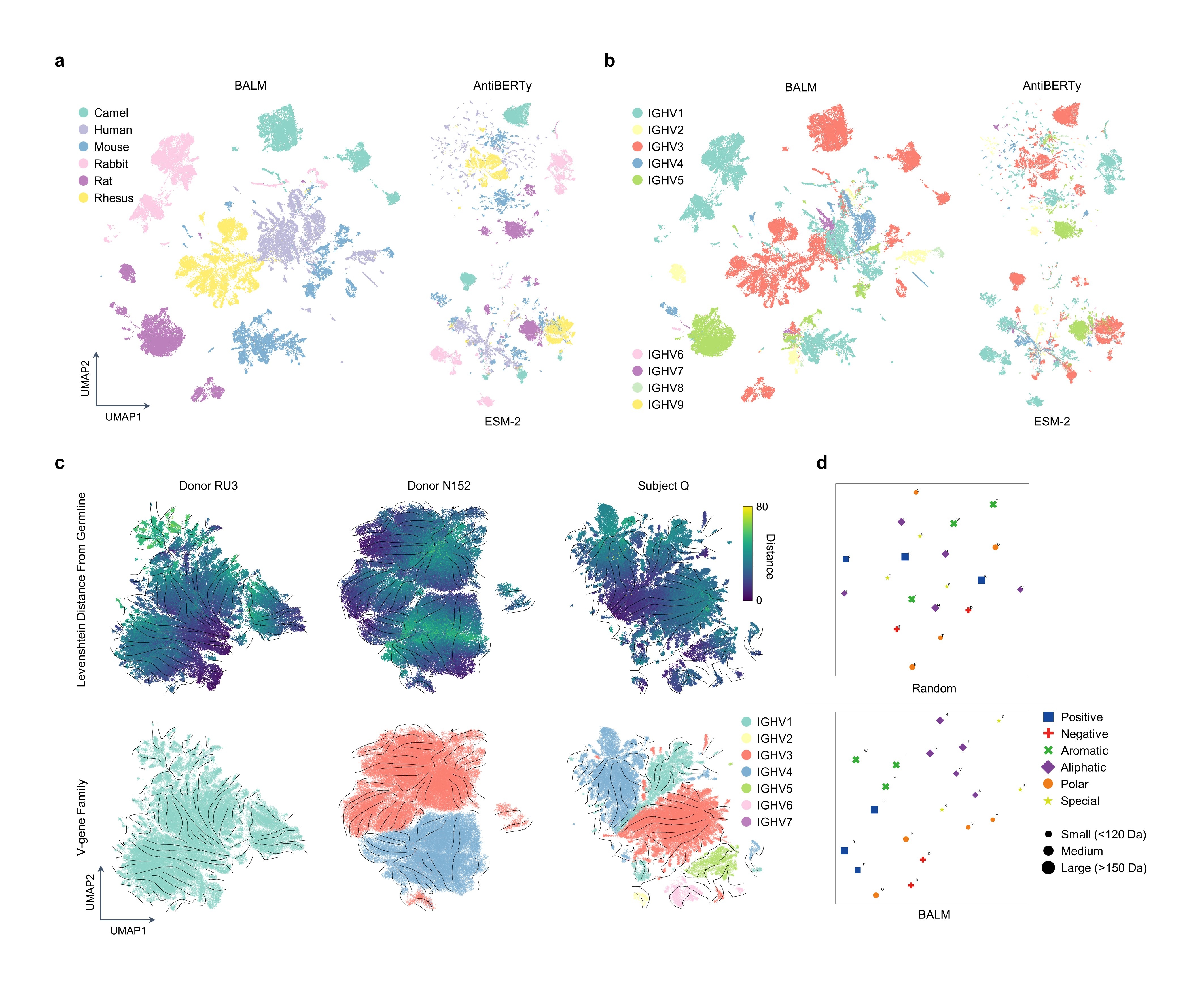}
\caption{\textbf{Representation of BALM encoding rich biological insights.}
    \textbf{a}, UMAP representation of species.
    A selection of 60,000 sequences from six species was evenly extracted from the OAS dataset. The final hidden layer of BALM was projected into a two-dimensional space using UMAP. The points are labeled according to diverse species, and the sequence embeddings from BALM are grouped by species. The delineation between various species is more pronounced compared to two other benchmark baselines.
    \textbf{b}, UMAP representation of V-gene family.
    Using the same dataset of 60,000 sequences, the points are labeled according to distinct V-gene families. The representation of BALM reveals clustering within the V-gene families.
    \textbf{c}, UMAP representation of evolutionary velocity. Arrows denote the direction of evolutionary velocity. In the top figures, points are annotated based on the Levenshtein distance from corresponding germlines to sequences. In the lower figures, patients with HIV and SARS-CoV-2, exhibiting different V-gene family counts, are shown. Donor RU3 and N152 are HIV patients predominantly exhibiting one and two V-gene families, respectively, while Subject Q is a SARS-CoV-2 patient with multiple V-gene families.
    \textbf{d}, t-SNE mapping of amino acid biochemical properties. The last hidden layer embeddings of both pre-trained (lower) and untrained (upper) BALM are projected into a two-dimensional plane using t-SNE. Each point represents an amino acid labeled with its biochemical properties. For the pre-trained BALM, residues cluster according to properties such as charge, aromaticity, and aliphatic nature. In contrast, the no-pretrain embedding space does not showcase fine-grained discrepancies in biophysical attributes.
}
\label{fig2}
\end{figure}

\textbf{BALM learns biological representation from unlabeled antibody sequences.}
Uncovering inherent patterns and properties is crucial for comprehending the specificity and function of antibodies. Recent evidence suggests that language models possess a potent capability for capturing semantic information encoded within input amino acid sequences, thereby facilitating the prediction of structure and function \cite{esmfold,antiberta,Prottrans}. 
By visualizing the embeddings of pre-trained language model, we are able to gain significant insights into the learned features of BALM and comprehend how it utilizes this information for making predictions across a variety of downstream tasks. For the purpose of examining the representation gleaned by BALM during its pre-training phase, we conducted a projection of 60,000 antibody sequences from the OAS dataset, selected randomly but with an equal number of sequences representing each species. We utilized the uniform manifold approximation and projection (UMAP) algorithm \cite{umap} to map the 640-dimensional final layer embeddings of these sequences into a two-dimensional space (Figs. \ref{fig2}a-\ref{fig2}b). Despite the absence of additional biological information besides antibody numbering, the acquired representations effectively categorize species and V-gene family, outperforming the models including ESM-2 and AntiBERTy. Particularly, BALM demonstrates superior capacity in differentiating human and mouse sequences compared to these two baseline models. Regarding variable gene families, the model might encounter challenges in their differentiation due to the insufficiency of sequence data available for each V gene family. Despite this, BALM's overall embedding is still reasonably proficient at grouping V-gene families.  All these suggest that the sequence alignment position embedding might provide some critical insight during the pre-training of BALM for identifying interesting patterns.

Antibodies possess amino acid composition and distribution that  share some similarities with common proteins, yet they exhibit unique biochemical properties owing to their specialized function in the immune system. The presence of charged, hydrophobic, aromatic, and polar residues contributes to the overall structure, stability, and function of antibodies, particularly in the CDRs crucial for antigen recognition. The 640-dimensional final hidden layer embedding of BALM is projected into two dimensions using t-SNE \cite{t-sne}.  When compared to random weights, pre-trained BALM reveals distinct clustering patterns of  antibody residues based on different biochemical properties (Fig. \ref{fig2}d).

\textbf{BALM learns antibody mutation trajectories from germline.}
Pharmaceutical development necessitates comprehensive insights into mutation trajectories and immune responses. Diverse repertoires arise through the process of somatic recombination, containing various stages of affinity maturation with distinct trajectories. Understanding the dynamics of immune responses for specific diseases aids in elucidating the orientation of the antibody affinity maturation and facilitates vaccine design. During B lymphocyte development, VDJ gene segments are randomly selected to create a functional VDJ exon. Different V gene segments, belonging to distinct families, encode the variable region of the heavy chain, resulting in a vast diversity of antigen-binding sites. In the adaptive immune system, V-gene segments contribute to varying levels of affinity maturation and specificity for foreign substances.

Taking Inspiration from the protein evolution analysis in \cite{evolocity}, the observation of locally optimal traversal in the antibody landscape can provide an intuitive explanation for  mutation strategies. Especially when dealing with multiple positional mutations beyond germline, language models learn broader high-dimensional representations of antibody sequences rather than individual changes. To construct a visual representation of mutation trajectories derived from germline sequences, we have collated three complete immune repertoires from individuals with the IGHG isotype. The repertoires of RU3 \cite{RU3} and N152 \cite{N152} consist of 77,271 and 102,213 HIV-specific unique sequences, respectively, while that of subject-Q \cite{subjectQ} comprises 60,026 unique sequences targeted against SARS-CoV-2.  By projecting the last hidden layer from the language model of sequences into two dimensions using UMAP (Fig. \ref{fig2}c), each node in the graph represents one antibody sequence.  The sequences are then connected with nearby sequences using the $k$-nearest neighbors (KNN) method. The shade of nodes depicts the Levenshtein distance between sequences and germline sequences, while the direction of arrows illustrates the affinity maturation trajectories of immune repertoires in different V-gene groups.

The language model uncovers the manifold features of affinity maturation for individual V-gene segments. Leveraging the inherent properties of antibodies, the language model effectively learns about variable regions and the distinct contributions of various V-gene segments to antibody responses. Despite the expectation of mutations favoring higher likelihood sequences in immune responses, we observe a directionality of evolution towards sequences that are closer to germline sequences for each cluster. The trajectory of sequences belonging to the same V-gene group reveals the affinity maturation of antibodies. The direction of evolutionary velocity indicates that the pre-trained language model tends to predict missing residues closer to low edit distance from germline rather than highly mutated sequences. Over time, as antibodies diverge from the original germline sequence, repertoires contain a greater number of sequences that are close to the germline during the affinity maturation process (see Supplementary Fig. \ref{LD}).

\begin{table}[h]
\centering
\caption{\textbf{Comparison of language models in predicting antibody binding characteristics.}
The mean values and the standard deviations for four assessment metrics are presented, with the standard deviation computed after three repetitions under identical configurations, excluding random seed variations. ESM-2 (containing 150M parameters) and ESM-1b (with 650M parameters) are included for comparison. The performance metrics of EATLM on antigen binding are derived from \cite{EATLM}, and average values of ProtBERT and AntiBERTa for paratope prediction are sourced from \cite{antiberta}. `BALM w/o PT' refers to BALM without the pre-training phase. Despite not being explicitly trained on antibody sequences and residue classification tasks, BALM achieves state-of-the-art performance in all evaluation metrics across the three tasks, except for the F1 score in the binding prediction task.
}
\label{table1}
\begin{center}
\begin{minipage}{\textwidth}
\centering
\begin{tabular}{llcccc}
  \hline
  Task      & Model  & F1 & MCC & AUC & APR \\\hline
  \multirow{7}{*}{Binding}  
  & BALM w/o PT    
  & $85.9\pm0.3$ & $70.7\pm0.3$ & $92.3\pm0.1$ & $92.6\pm0.2$
  \\
  & AntiBERTy     
  & $85.5\pm0.4$ & $68.9\pm1.0$ & $92.1\pm0.0$ & $92.7\pm0.1$ 
  \\ 
  & AbLang-H     
  & $86.1\pm0.1$ & $70.2\pm0.4$ & $92.3\pm0.1$ & $92.3\pm0.3$ 
  \\ 
  & EATLM 
  & $\textbf{86.2}\pm\textbf{0.4}$ & $69.9\pm1.0$ & $92.2\pm0.4$ & $ - $
  \\ 
  & ESM-2
  & $85.5\pm0.2$ & $69.1\pm0.3$ & $92.1\pm0.2$ & $92.7\pm0.1$
  \\
  & ESM-1b
  & $85.4\pm0.1$ & $69.4\pm0.1$ & $92.1\pm0.1$ & $92.6\pm0.1$
  \\
  & \textbf{BALM}
  & $\textbf{86.2} \pm \textbf{0.2}$ & $\textbf{70.9} \pm \textbf{0.6}$ & $\textbf{92.4}\pm\textbf{0.1}$ & $\textbf{92.9}\pm\textbf{0.1}$
  \\
  \hline
  \multirow{7}{*}{Paratope}  
  & BALM w/o PT     
  & $66.0\pm1.0$ & $62.5\pm1.1$ & $96.0\pm0.1$ & $71.4\pm0.7$
  \\ 
  & ProtBERT      
  & $68.6\pm0.0$ & $65.2\pm0.0$ & $95.9\pm0.0$ & $70.1\pm0.0$
  \\ 
  & AntiBERTa      
  & $68.9\pm0.0$ & $65.9\pm0.0$ & $96.1\pm0.0$ & $74.2\pm0.0$
  \\ 
  & AntiBERTy     
  & $67.9\pm1.3$ & $64.6\pm1.4$ & $95.7\pm0.1$ & $72.8\pm1.0$
  \\ 
  & AbLang-H     
  & $64.8\pm7.6$ & $62.6\pm5.8$ & $95.6\pm0.3$ & $72.4\pm1.0$ 
  \\ 
  & ESM-2     
  & $68.0\pm1.1$ & $64.7\pm1.0$ & $\textbf{96.3}\pm\textbf{0.1}$ & $74.2\pm0.4$
  \\ 
  & ESM-1b     
  & $69.4\pm0.4$ & $66.2\pm0.5$ & $95.7\pm0.7$ & $73.0\pm0.8$ 
  \\ 
  & \textbf{BALM}   
  & $\textbf{70.1}\pm\textbf{0.8}$ & $\textbf{67.0}\pm\textbf{0.9}$ & $\textbf{96.3}\pm\textbf{0.2}$ & $\textbf{76.2}\pm\textbf{0.1}$ 
  \\ 
  \hline
  \multirow{7}{*}{Redundancy}
  & BALM w/o PT
  & $59.0\pm3.2$ & $13.6\pm1.6$ & $59.8\pm1.0$ & $57.1\pm1.1$
  \\
  & AntiBERTy
  & $61.5 \pm 1.4$ & $23.4\pm0.8$ & $66.9\pm0.2$ & $64.1\pm0.2$
  \\
  & AbLang-H
  & $64.5\pm0.8$ & $23.4\pm0.4$ & $68.5\pm0.2$ & $66.0\pm0.2$
  \\
  & ESM-2
  & $52.1\pm3.6$ & $15.1\pm1.2$ & $62.0\pm0.7$ & $59.5\pm0.2$
  \\
  & ESM-1b
  & $51.7\pm2.0$ & $17.3\pm0.8$ & $63.9\pm0.5$ & $61.0\pm0.6$
  \\
  & \textbf{BALM}
  & $\textbf{67.1}\pm\textbf{0.7}$ & $\textbf{34.7}\pm\textbf{1.0}$ & $\textbf{75.6}\pm\textbf{0.2}$ & $\textbf{76.1}\pm\textbf{0.5}$
  \\
  \hline
\end{tabular}

\end{minipage}
\end{center}
\end{table}

\textbf{BALM learns binding properties from large-scale antibody sequences.}
BALM was initially pre-trained on a vast collection of unlabeled sequences, employing a masked language model self-supervised training task. Subsequently,  fine-tuning was performed on a variety of antibody domain tasks to assess the effectiveness of the learned representation. The downstream functional tasks include predicting and characterizing multiple facets of antibody function. Four specific tasks were evaluated for BALM, addressing essential questions related to antibodies: 1) Antigen-binding capacity \cite{Mason2021, EATLM}; 2) Binding site locations \cite{antiberta}; 3) Redundancy in immune responses to antigens \cite{newOAS}; and 4) Antigen affinity \cite{SKEMPI,S1131,M1707}. Gaining insights into target antigen binding patterns and discerning desirable therapeutic properties can accelerate the drug discovery process. However, identifying antibodies requires numerous labor-intensive experimental methods. Rapid and precise prediction of specificity and affinity to target molecules is essential for developing novel therapies and understanding the immune system. Evaluation metrics include Matthews' correlation coefficient (MCC), area under the receiver operating characteristic curve (AUC), and average precision scores with recall (APR).

Optimizing therapeutic antibodies necessitates determining their binding specificity with target antigens. The prediction of antigen binding is a binary sequence-level classification task that involves interactions with binding targets, where the CDR H3 region plays a pivotal role. Gaining insights into binding site characteristics and predicting antigen-specific variants offer considerable advantages for clinical pharmaceutical development and antibody engineering. To investigate the interaction between the target antigen human epidermal growth factor receptor 2 (HER2) and the clinically approved wild-type antibody trastuzumab, we compiled a dataset of antibody-expressing sequences that replace the wild-type trastuzumab sequence with variant heavy chain CDR H3 fragments, as obtained from \cite{Mason2021, EATLM}. The antigen-binding dataset, derived from one germline sequence, comprises 21,612 sequences and follows a 70\%/15\%/15\% split. Despite the high similarity in amino acid composition per position between binding and non-binding sequences, BALM achieves an APR of 92.9 and outperforms other language models across nearly all metrics. 
Remarkably, even without pre-training, BALM leverages inherent antibody features and delivers results comparable to other pre-trained models (Table \ref{table1}), showcasing its robust predictive capabilities in specificity.

\begin{figure}[h!]
	\centering
    \includegraphics[width=\linewidth]{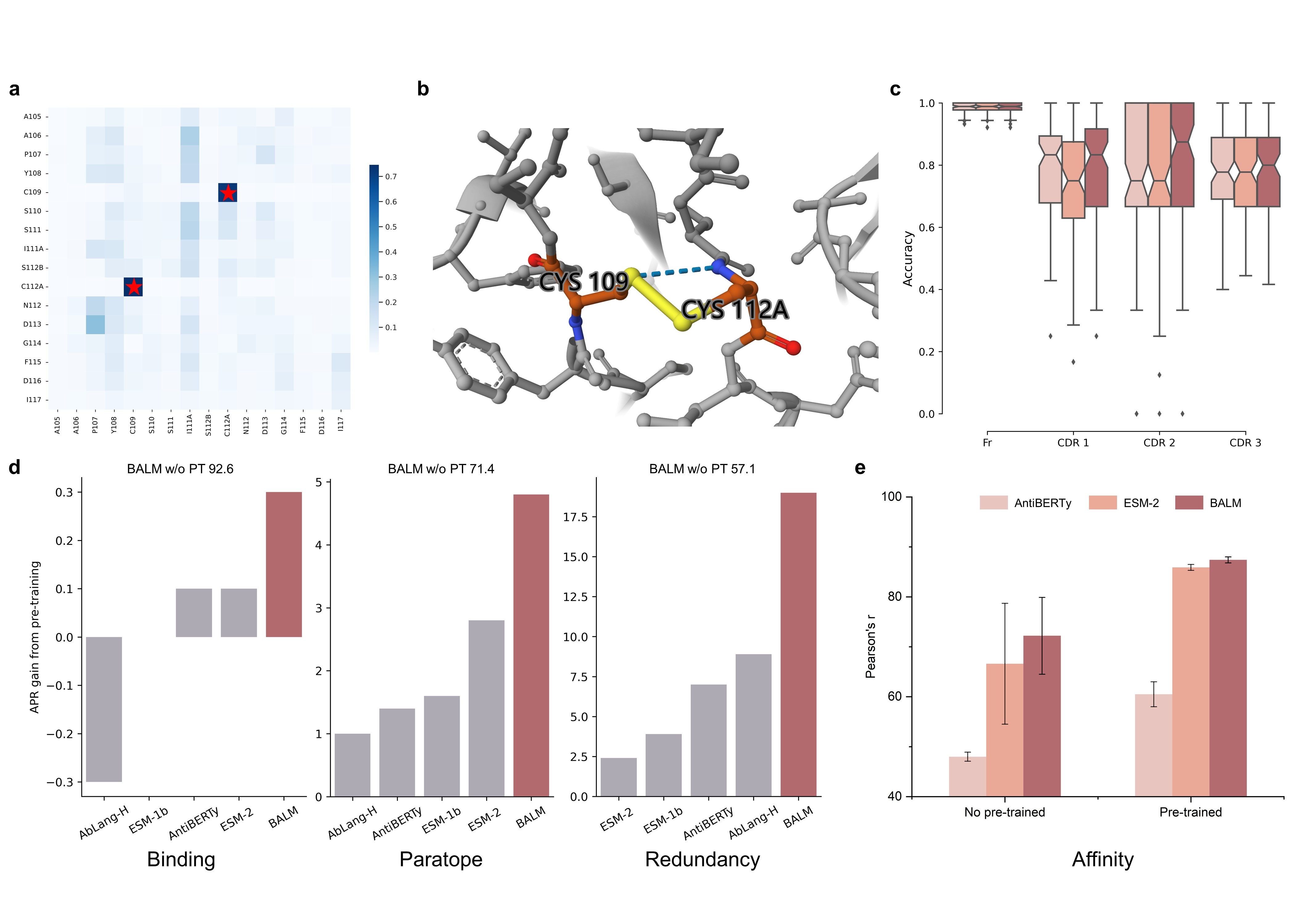}
\caption{\textbf{Efficient enhancement of antibody engineering by BALM.}
\textbf{a}, Self-attention map for the CDR H3 of the SARS-CoV-2 therapeutic antibody Tixagevimab's heavy chain, derived from the last hidden layer of the $11^{\rm th}$ head in BALM. The horizontal and vertical axes represent the residues in the CDR H3 and the positions according to IMGT numbering, respectively. More intense cell coloration indicates higher attention weights and stronger contextual correlations between the associated residues.
\textbf{b}, A crystallographic structure (PDB: 8D8R) depicting the interactions mediated by hydrogen and disulfide bonds between two cysteine residues at positions 109 and 112A in Tixagevimab's heavy chain, as illustrated in the left figure.
\textbf{c}, Paratope prediction across CDRs and framework.
\textbf{d}, APR gain from pre-training across five selected methods compared with baseline BALM w/o pre-training.
\textbf{e}, Comparison of affinity maturation prediction, assessed through Pearson's $r$, with other language models. Each error bar represents the standard deviation. AntiBERTy with 26M parameters were fine-tuned to compare.
 }
\label{fig3}
\end{figure}

Paratopes, the antibody binding sites that interact with antigens, are critical for understanding the binding mechanism and optimizing antibody binding properties in structural immunology. Predicting paratopes involves a token binary classification task, where the binding probability is determined for each position. By employing the antibody numbering scheme and emphasizing on CDRs modeling, BALM can take advantage of additional sequence region categorization information.  This enables accurate prediction of key residues, facilitating the optimization of CDRs. Moreover, delving into the binding mechanism for each position of CDR and non-CDR regions is crucial, extending beyond the recognition of CDR 3, which predominantly governs antibody function. BALM demonstrates superior performance compared to existing methods and outperforms AntiBERTa by 2 points on APR. Comparing the accuracy across various regions, BALM consistently outperforms the two baseline models on all CDRs, as depicted in Fig. \ref{fig3}c. Interestingly, even without pre-training, BALM with evolutionary position embedding achieves superior performance compared to the protein language model ProtBERT \cite{Prottrans}.

Antibody redundancy, characterized by the capacity of multiple antibodies to recognize and bind to the same antigen, is a significant facet of our immune response. During somatic hypermutation (SHM), antibody genes can undergo a plethora of mutations, engendering a broad range of antibodies with varying antigen specificities. This variability enhances the immune system's ability to identify and eliminate pathogens. High redundancy in neutralizing antibody responses represents a robust mechanism to withstand mutations, underscoring redundancy's importance in enabling the immune system to efficiently neutralize a diverse array of antigens \cite{barabasi2004network}. Understanding antibody redundancy can guide the development of therapeutic approaches that deploy the most potent antibodies to recognize specific antigens, ensuring resistance to degradation or denaturation across various environments. To assess the performance of BALM in contrast to other pre-existing pre-trained language models, we established a classification task focusing on antibody redundancy and its varying degrees of robustness against SARS-CoV-2. After applying the preprocessing procedures outlined in Methods Section, we derived a final dataset comprising 31,718 sequences labeled as high or low. Remarkably, our BALM model excelled in all metrics, even though the language model training paradigm did not incorporate any additional labels beyond masked residues. BALM achieved an impressive APR score of 76.1, outpacing the second-best model by a margin of 10 points (Table \ref{table1}). Even without pre-training, BALM with randomly assigned weights demonstrated comparable performance to ESM-2. The integration of additional biological features into an antibody language model enhanced its performance, surpassing the outcomes of standard pre-training processes. In addition, the augmentation in APR scores attributed to pre-training intensifies as tasks become more tightly linked with antibodies (Fig. \ref{fig3}d).

The binding affinity of an antibody is a multifaceted attribute, influenced by the antibody's variable regions, epitope external features, and environmental determinants. Elevated affinity is indicative of enhanced antigen neutralization and initiation of downstream immune responses. In vitro experiments designed for affinity maturation can augment the binding capability of an antibody against its target antigen. The alignment of antibody sequences streamlines the analysis of mutations, thereby enhancing the comparability of SHM in B cells. These cells produce antibodies exhibiting heightened affinity towards specific antigens. The binding free-energy change ($\Delta \Delta G$) quantifies the stability alterations resulting from mutations in protein sequences, comparing wild-type (WT) and mutation-type (MT) sequences. This value embodies the free energy discrepancy between the bound and unbound states of protein-ligand complexes. By utilizing the pre-trained antibody-specific BALM, we can precisely and efficiently predict binding affinity variations of residue mutations against multiple SARS-CoV-2 variants. Our binding affinity dataset, an antibody subset derived from the structural kinetic and energetic database of mutant protein interactions (SKEMPI) V2.0 \cite{SKEMPI}, encompasses experimentally determined binding free energy changes upon mutation for protein-protein interactions. We computed Pearson's $r$ between the experimental and predicted $\Delta \Delta G$ values using our model and other benchmark models. As depicted in Fig. \ref{fig3}e, BALM's performance surpasses that of ESM-2 and AntiBERTy in terms of Pearson's $r$ score. Notably, without pre-training, BALM achieved an average score of 72.2, outpacing the scores of ESM-2 by 5.6 and pre-trained AntiBERTy by 11.7. When compared to BERT-based models with a similar parameter count such as BALM and ESM-2, our BALM model equipped with antibody-specific positional embedding achieves a superior Pearson's $r$ score.

\textbf{BALM encodes antibody structural representations.} The relative paucity of experimentally validated antibody structures (approximately a thousand), presents a substantial impediment for the precise prediction of antibody structures. BALM distils patterns from extensive sequence datasets to counterbalance this structural information deficit. The multi-head self-attention mechanisms integral to the pre-trained antibody language model are adept at capturing the sophisticated interdependencies among various positions within the antibody molecule. As a practical demonstration of BALM's proficiency in learning pairwise interaction patterns, we selected to use Tixagevimab as the example input, a recently approved SARS-CoV-2 therapeutic monoclonal antibody. Figure~\ref{fig3}a exhibits the $11^{\rm th}$ head of BALM's last hidden layer, with high attention scores serving as indicators of significant contextual associations among the residues. The heatmap of CDR H3 reveals a pair of cysteine residues at positions 109 and 112A, distinguishable by their darkest hue. Correspondingly, the crystal structure presented in Fig. \ref{fig3}b shows that these cysteine residues are linked via two non-covalent interactions, specifically hydrogen and disulfide bonds. The heatmap's inferred contacts offer an intuitive approach for implicitly defining antibody structure. Significantly, BALM exhibits the capacity to discern the underlying information embedded in Tixagevimab's heavy chain, making it valuable for  understanding the three-dimensional structure of the antibody.

\textbf{BALMFold accurately and efficiently predicts on antibody structure benchmark with single sequences.}
Although AlphaFold2 has attained significant accuracy in predicting protein structures, a reliable antibody structure prediction still presents a conundrum due to high variability and the scarcity of potent MSAs information. The capacity of antibodies to bind to antigens is predominantly influenced by the structural attributes of CDR loops. The pronounced variability that characterizes these CDRs serves to augment their functional versatility, yet concurrently introduces a layer of intricacy to any attempts at structural forecasting. For predicting antibody structures from sequences, the language model emerges as a critical element in extracting sequence embedding. BALM has earlier been shown to successfully encapsulate the inherent biological characteristics of antibodies. Derived from BALM, we introduce BALMFold, a novel tool for the precise prediction of full-atom antibody structures, with a particular focus on CDR loops. The performance evaluation of BALMFold was carried out using the same benchmark dataset as IgFold, comprising 197 paired and 71 nano antibodies from SAbDab \cite{SAbDab} spanning July 2021 to September 2022. It is important to note that the training set shares no similarity cutoff with the benchmark dataset, which was made public after the cutoff date in July 2021. BALMFold was compared to a series of deep learning-based structure prediction methods \cite{igfold,alphafold2,afm,omegafold,esmfold}.

\begin{figure}[H]
    \centering
\includegraphics[width=\linewidth,height=0.95\linewidth]{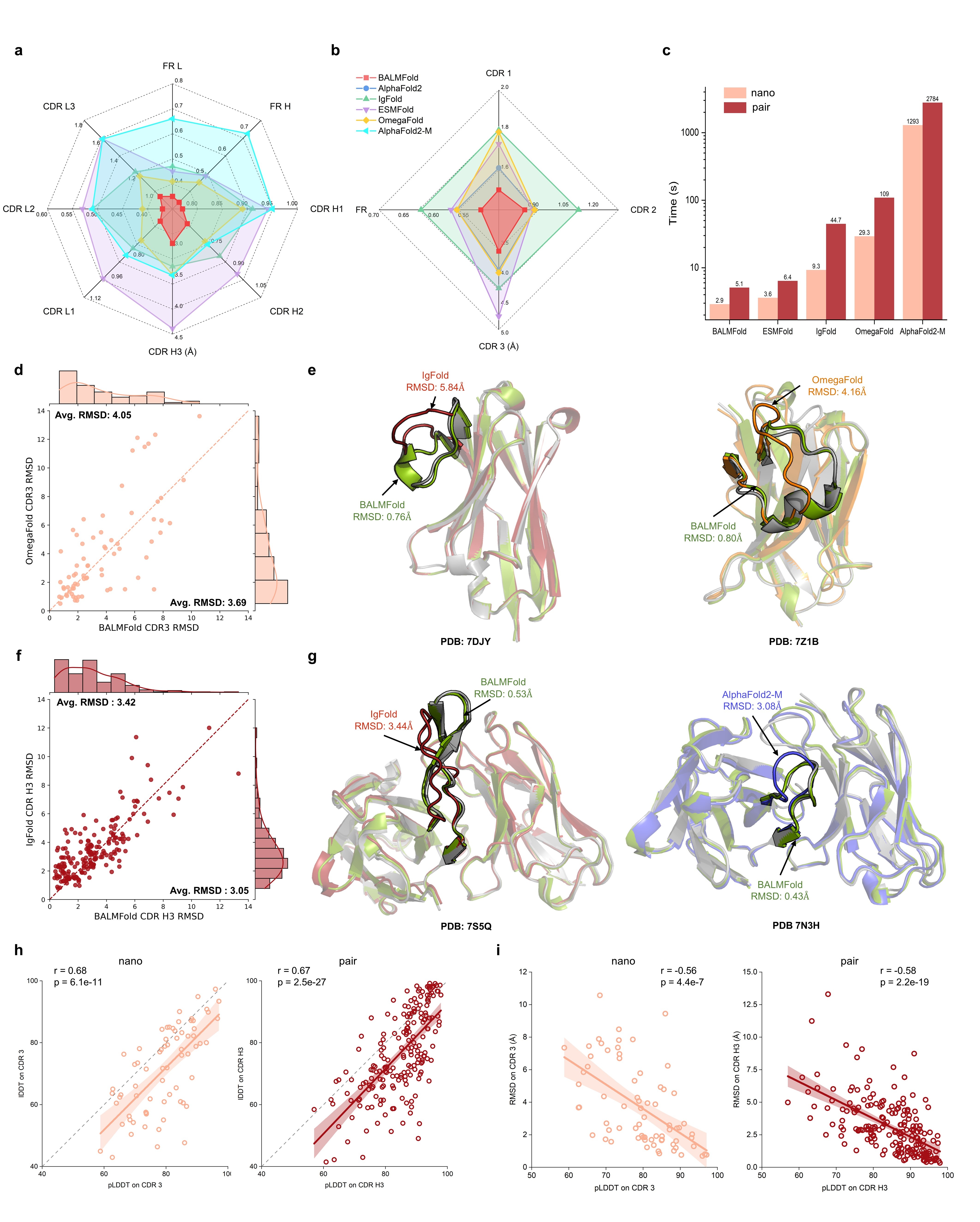}
    \caption{\textbf{BALMFold accurately predicts antibody structure from sequences.}
    \textbf{a}, Mean RMSD performance on the IgFold benchmark for 197 paired antibodies, considering eight regions. The label AlphaFold2-M refers to the multimer variant of AlphaFold2. Higher structure prediction accuracy is correlated with \textbf{smaller} model areas depicted in the figure, reflecting lower RMSD values.
    \textbf{b}, Mean RMSD performance on the IgFold benchmark for 71 nanobodies, spanning four regions.
    \textbf{c}, Comparative analysis of average structure prediction runtimes.
    \textbf{d}, Head-to-head comparison of RMSD values for the CDR H3 loop between BALMFold and OmegaFold for nanobodies. Predicted targets by BALMFold falling into the upper-left quadrant indicate lower RMSD than alternate method.
    \textbf{e}, Visualization of nanobody native experimental structure (grey) and predictions by BALMFold (green), IgFold (red), and OmegaFold (orange) for target 7DJY ($\mathcal{L}_\text{CDR H3}=14$ residues) and target 7Z1B ($\mathcal{L}_\text{CDR H3}=18$ residues) respectively. 
    The CDR H3 loops are highlighted. 
    \textbf{f}, Head-to-head comparison of RMSD values for the CDR H3 loop between BALMFold and IgFold for paired antibodies.
    \textbf{g}, Visualization of paired antibody native experimental structure (grey) and predictions by BALMFold (green), IgFold (red), and AlphaFold2-M (blue) for target 7S5Q ($\mathcal{L}_\text{CDR H3}=20$ residues) and target 7S8H ($\mathcal{L}_\text{CDR H3}=13$ residues) respectively.
    \textbf{h}, Illustration of the correlation between pLDDT and lDDT for $C_{\alpha}$ atom, with Pearson's $r$ and two-sided t-test P-value provided. The shaded region corresponds to the 95\% confidence interval.
    \textbf{i}, Scatter plots with regression lines between pLDDT and atomic RMSD values.
    }
\label{fig4}
\end{figure}

We conducted the evaluation of root-mean-square deviation (RMSD) values in relation to structures determined through experimental methods. Following an optimal superimposition process conducted with the corresponding chain of experimental structures, the RMSD values were separately calculated in each regions according to Chothia numbering. The calculation focused on all the backbone heavy atoms present in CDR loops and frameworks. As evidenced in Figs. \ref{fig4}a-\ref{fig4}b, the \textbf{smaller} regions observed in the radar graph correspond to lower RMSD values, indicating superior performance. The average RMSD values of BALMFold significantly outperform all other approaches across all CDRs and frameworks for both paired and single-chain antibodies. In addition, the orientational coordinate distance (OCD) \cite{ocd} is employed to assess the orientation of the heavy and light chains in paired antibodies (in Supplementary Table \ref{table: pair benchamrk}). In an investigation of 197 paired antibodies, BALMFold achieved a superior average RMSD score $\sim\SI{3}{\angstrom}$ specifically on the CDR H3 loop (Supplementary Table \ref{table: pair benchamrk}). Meanwhile, for the additional set of 71 nanobodies, BALMFold accomplished an average value of $\SI{3.69}{\angstrom}$ on the CDR 3 loop, contrasting with other four models which exceeded $\SI{4}{\angstrom}$ (Supplementary Table \ref{table: nano benchamrk}). Without explicitly considering inter-chain interactions, our method surpass AlphaFold2-Multimer with an average RMSD of $< \SI{0.51}{\angstrom}$ on CDR H3. Unlike AlphaFold2, BALMFold avoids reliance on the laborious process of searching for MSAs and templates, harnessing the pre-trained BALM to extract structural and evolutionary features directly from sequences. In comparison to the methods featuring a similar structure module to AlphaFold2, our language model exhibits noteworthy prowess in antibody structure prediction solely from antibody sequences. Balancing high-quality predictions for both nanobodies and paired antibodies is a common challenge among existing structure prediction approaches. While IgFold and the multimer version of AlphaFold2 show superior performance with paired antibodies, they falter with nanobodies. Notably, we find that amino acid sequences augmented with additional biological subregion representations allow BALMFold to accurately predict structures for both antibody types.

Computational efficiency is paramount for wide-scale practical applications in the design of antibody therapeutics. On average, BALMFold can predict antibody structures from amino acid sequences within 5 seconds. In contrast to AlphaFold2 and AlphaFold2-Multimer, which heavily lean on the costly procedure of searching for MSAs and templates, BALMFold employs BALM to extract evolutionary signals, circumventing the need for additional MSAs or templates. This provides for swift and precise antibody structure prediction. For a runtime comparison, the models were run on a single NVIDIA V100 GPU. As shown in Fig. \ref{fig4}c, BALMFold's inference speed significantly outstrips that of both AlphaFold2 and AlphaFold2-Multimer, is approximately six times faster than IgFold, and is comparable to ESMFold.

\textbf{Accurate prediction on CDR 3 loop.}
The extreme variability inherent to the structure and sequence of the CDR 3 loop imparts antibodies with a profound capacity to recognize and bind to an expansive range of antigens. This diversity underscores the CDR 3 as the primary determinant of the vast repertoire of antibodies. Nevertheless, this characteristic also complicates the accurate prediction of the CDR H3 loop, presenting a significant challenge in the field.

While five of the CDR loops conform to relatively predictable canonical folds, our contributions extend to all CDR loops with a particular emphasis on the CDR H3 loop. Utilizing the pre-trained BALM, our approach predicted the CDR H3 loop within $\SI{2}{\angstrom}$ RMSD for 70 out of 197 targets in the paired antibodies benchmark, reflecting a 49\% improvement over the next best performing method. Notably, our method BALMFold achieved a minimum RMSD of $\SI{0.34}{\angstrom}$ on target 7UEN. Furthermore, BALMFold's predictions consistently demonstrate sub-$\SI{1}{\angstrom}$ RSMD on the CDR 3 loop of the light chain. Within the single-chain antibody benchmark, approximately 35\% of BALMFold's predicted targets outperform the other four models by achieving the lowest RMSD. BALMFold notably exhibited a superior performance on hard targets as well, producing an RMSD of $\SI{2.4}{\angstrom}$ on 7M1H, while alternative models surpassed $\SI{5}{\angstrom}$.

As expected, increasing the length of the CDR 3 loop resulted in greater conformational diversity, thus compounding the prediction difficulty. We observed a correlation describing how the RMSD of prediction rises monotonically with the length of the CDR3, with paired antibodies displaying heightened sensitivity to the length of the CDR3 compared to single-chain antibodies (Supplementary Fig. \ref{figure: rmsd vs cdr3 length}). We performed direct comparisons of BALMFold with two competing models, OmegaFold and IgFold, across two domain datasets (Figs. \ref{fig4}d-\ref{fig4}f). Notably, scatter plots indicate superior performance of BALMFold in upper triangle regions. As depicted in Fig. \ref{fig4}d, BALMFold surpassed OmegaFold by achieving lower average RMSD values and standard deviation. Against IgFold, BALMFold outperformed by $\SI{0.37}{\angstrom}$ with P-value of 0.07 in a two-sided $t$-test on paired antibodies (Fig. \ref{fig4}f).

To further illustrate the robustness of our model, we conducted case studies on two domains to evaluate accurate prediction of the CDR loop (Figs. \ref{fig4}e-\ref{fig4}g). For single-chain instance 7DJY, BALMFold demonstrated superior accuracy, vastly outperforming IgFold ($\SI{0.76}{\angstrom}$ versus $\SI{5.84}{\angstrom}$).  Clearly, the latter's predictions deviated entirely from the native conformation, while BALMFold presented divergent predictions showcasing markedly-improved accuracy in loop conformation predictions previously mispredicted by IgFold.
In the case of paired antibody target 7N3H, BALMFold achieved a significantly lower prediction RMSD ($\SI{0.43}{\angstrom}$) than AlphaFold2-M ($\SI{3.08}{\angstrom}$).

To estimate the predictive reliability of BALMFold, we computed the local distance difference test (lDDT) for each residue's $C_{\alpha}$ atom. For high diversity CDR 3, we observed a notable correlation between the predicted lDDT (pLDDT) and the actual lDDT (Fig. \ref{fig4}h). This strong correlation is reflected by Pearson's $r$ of 0.68 for nanobodies and 0.67 for paired antibodies. Regression plots for CDR 1 and CDR 2 are provided in Supplementary Fig. \ref{figure: plddt with lddt}. In terms of the confidence level for the predicted residue-residue distances and structural precision, we conducted a further exploration of the correlation between pLDDT and RMSD values (Fig. \ref{fig4}i). We observed a close correlation, marked by Pearson's $r$ values of -0.56 for nanobodies and -0.58 for paired antibodies. The plots for other CDRs are depicted in Supplementary Fig. \ref{figure: plddt with rmsd}. These observations provide crucial perspectives in identifying probable regions of unreliable prediction and assess the overall quality of the predicted antibody structures.

The comprehensive evaluation of BALMFold underscores its superiority over contemporary methods in accurately predicting the structure of both antibodies and nanobodies. Specifically, in addressing long-standing challenges associated with CDR 3 loop modeling, our method exhibits significant advancements, bolstering its potential impact on the field.

\section{Discussion}\label{sec12}
Language models have demonstrated their capacity to learn inherent biological information from sequences, enabling them to predict functional properties and structures of antibodies \cite{esm1b,esmfold,antiberta,Antiberty}. Anfinsen's dogma posits that the native conformation of a protein is exclusively determined by its amino acid sequence \cite{anfinsen}. Antibody-specific tasks often lack effective MSAs and templates for CDRs. Moreover, 40\% of antibody repertoires in the OAS dataset are missing the first 15 amino acids \cite{ablang}. To thoroughly consider the unique complementarity-determining regions and framework regions of antibody sequences, we propose a bio-inspired antibody language model trained on 336M antibody sequences, which has proven its effectiveness in prediction of various antigen functional properties. We conducted an extensive analysis to uncover the evolutionary trajectories of antibody sequences in response to specific diseases. When combined with a language model, several machine learning approaches have demonstrated their success in predicting protein structures \cite{alphafold2,esmfold,omegafold,afm,RGN2,igfold}. Capitalizing on the meaningful representations of language models, BALMFold is developed to predict structures directly from antibody sequences. Exploiting the biological features of antibodies, BALMFold outperforms IgFold, OmegaFold, AlphaFold2, and ESMFold on benchmark that contains 268 antibodies. By eliminating the search process for MSAs, BALMFold predicts structures more rapidly than MSA-based and template-based approaches. Accurate and fast prediction of antibody structures is crucial for understanding the interactions between antibodies and their target antigens, even without explicit evolutionary information.

While BALM has achieved impressive results across various tasks, antibody structure prediction with atomic accuracy remains an unsolved challenge, particularly for CDR 3. Considering inter-chain interactions could enhance the predicted accuracy of paired antibodies. Drawing inspiration from existing numbering schemes, specific antibody numbering schemes for language models could be developed in the future. Limited by computational resource, higher capacity language model potentially yield better performance. By incorporating bio-inspired antibody positional embedding and antibody features, analogous generative models after training hold the potential of designing therapeutics in antibody discovery \cite{shin2021protein}. Biologically motivated approach can effectively extract representations of antibody sequences to advance antibody research and therapeutic development.

\section{Methods}\label{Methods}

\subsection{Overview of BALM}
To comprehend antibody functional properties and structures, we propose an antibody-specific language model designed to efficiently capture the rich information and representation inherent in repertoires. The language model's learned representation can provide biological properties essential for function and structure prediction in antibody engineering.

Largely adhering to the ESM-2 architecture \cite{esmfold} comprising 150 million parameters, we incorporated 30 transformer encoder blocks \cite{transformer}, each containing 20 multi-head self-attention layers and a feed-forward layer with 640 hidden states, layer normalization, and residual connections. The beginning of each sequence is combined with a classification token, enabling the prediction of various tasks. In consideration of the maximum length of the variable heavy chain region of antibodies, input sequences are padded or truncated to a standardized length of 168. While the original ESM-2 architecture employs rotary position embedding (RoPE) \cite{RoPE} to supply token positional information, we substituted RoPE with our bio-inspired antibody positional embedding to account for the distinct functional regions of antibody sequences.

\subsection{Bio-inspired Antibody Positional Embedding (BAPE)}
In the antibody-specific language model, each amino acid is regarded as a token and projected into token embedding. Due to the permutation invariant self-attention mechanism's lack of inherent consideration for the order of input tokens, positional embedding encodes the location information of tokens in the sequence, assigning each position a unique representation. This positional embedding is then added before being fed into the encoder block.

Existing alternative approaches, however, do not incorporate antibody evolutionary information. To effectively harness sequential information, we propose a bio-inspired antibody position embedding that aims to: (1) exploit evolutionary information in variable regions of antibody sequences, (2) reduce dependency on high-quality MSAs, and (3) address the deficiency of a complete antibody corpus in the OAS dataset.

The ImMunoGeneTics (IMGT) system \cite{imgt} is a standardized numbering method for annotating variable domains of immunoglobulins (IGs) and T cell receptors (TRs). The unique IMGT numbering relies on the highly conserved structural features of the variable region. Different regions are expected to contain a specific number of amino acids. Gaps are created in regions with fewer amino acids than expected, and additional positions are inserted between positions 111 and 112 for more than 13 amino acids in the CDR3. Consequently, there are 128 positions for IG V-domain annotation in antibody sequences with no more than 13 amino acids in the CDR3.

By specifically emphasizing the positions of CDRs and framework regions, we observe that BALM effectively investigates functionally important regions in a series of downstream tasks and structure prediction tasks. The unique numbering system enhances the prediction and understanding of sequences and structures across a diverse array of antibodies.

\subsection{BALMFold architecture}
Leveraging the potent capabilities of the antibody-specific model BALM, BALMFold demonstrates remarkable precision in predicting antibody structures, especially within the highly variable CDRs. The primary architecture of BALMFold comprises two elements: BALM, responsible for extracting information from the antibody sequence, and the folding block, inclusive of the BAformer and structure module, as illustrated in Fig. \ref{fig1}a.

Within the framework, BALM manipulates the antibody sequence to construct residue embeddings, thereby encoding amino acid representations that have been learned. Pair embeddings are generated by integrating singular representation pairwise embeddings, thus encoding learned residue-residue interactions. The subsequent step involves introducing these residue and pair embeddings into the BAformer, a four-layered structure which refines features via the exchange of singular and pair representations.

The BAformer is structured into two separate elements that individually update single and pair embeddings. The first component incorporates four self-attention and transition blocks, which together form a feed-forward network responsible for updating the single embedding by discerning global dependencies within antibody sequences. Following its update, the single embedding is subject to a pairwise product procedure, computing interactions between individual amino acids via an outer product operation. The second component merges the pair embedding with the computed pairwise features, and subsequent to this combination, the pair embedding is updated utilizing four triangle update and transition blocks. Consequently, a single layer of the BAFormer module aligns functionally with four layers of the embedding update modules in AlphaFold2 \cite{alphafold2}, but delivers greater computational efficiency. The structural module with shared weight parameters subsequently employs the refined residue and pair embeddings to predict the 3D coordinates of protein backbones and side chains. It comprises eight invariant point attention (IPA) layers are tasked with predicting the positions, orientations, and angles of backbones and side chains. Following each layer's completion, the projected positions are transferred to the succeeding layer to act as structural initiators. Furthermore, at a global level, we cycle single and pair embeddings back to the BAformer three times. With the support of our pre-trained BALM, our model negates the necessity for exhaustive searches for sequence homologs and structural templates, significantly diminishing runtime while sustaining excellent accuracy levels.

During the BALMFold training phase, we freeze the weights of our pre-trained antibody language model to minimize the loss function. As indicated in Eq.~\eqref{equation4}, the ensemble training loss combines four components: frame aligned point error (FAPE) loss on all atoms denoted by $\mathcal{L}_{\text {FAPE}}$, distogram loss $\mathcal{L}_{\text {dist}}$, confidence loss $\mathcal{L}_{\text {pLDDT}}$, and structure violation loss $\mathcal{L}_{\text {viol}}$.
The primary component $\mathcal{L}_{\text {FAPE}}$, introduced by AlphaFold2 \cite{alphafold2}, quantifies the discrepancies in inter-residue distances and orientations between predicted and actual atom coordinates following global alignment. 
Additionally, $\mathcal{L}_{\text {dist}}$ signifies an averaged cross-entropy loss aimed at minimizing the divergence between the predicted and actual distogram derived from antibody structures, and $\mathcal{L}_{\text {pLDDT}}$ involves the predicted lDDT \cite{lddt} extracted from the final residue representations of the structure module, intended to penalize instances where confidence and accuracy are misaligned. Finally, $\mathcal{L}_{\text {viol}}$ includes violations of steric constraints, such as angles and bond lengths.

\begin{equation}
\mathcal{L} =  \mathcal{L}_{\text {FAPE}}+ \mathcal{L}_{\text {dist}}+ 0.01\mathcal{L}_{\text {pLDDT}}+ 0.01\mathcal{L}_{\text {viol}}. 
\label{equation4}
\end{equation}

\subsection{Datasets}
\noindent \textbf{BALM training dataset}
The immune repertoires utilized for training are derived from the Observed Antibody Space (OAS) \cite{newOAS} dataset, encompassing over one billion annotated antibody sequences from more than 80 distinct studies. All unpaired antibody sequences in OAS, including heavy and light chains, were clustered at $40\%$ identity using LinClust \cite{LinClust} to eliminate redundant sequences. This process resulted in 359 million non-redundant sequences, comprising 358 million heavy chains and 1.5 million light chains. Data lacking the \textit{sequence\_alignment\_aa} feature were filtered out, and $1\%$ of the sequences were reserved for validation. Consequently, the final training dataset contains 336 million non-redundant sequences.

\noindent \textbf{BALMFold training dataset}
\label{antibody_structure_data}
We assembled a collection of experimentally determined antibody structures available prior to July 1st, 2021, featuring a resolution below $\SI{3}{\angstrom}$ from the SAbDab database\cite{SAbDab}. To cluster the antibody sequences, we utilized MMseqs2 \cite{Steinegger2017MMseqs2ES} at a $99\%$ sequence identity threshold. Redundant structures within the same cluster were removed. The final training dataset encompasses 2371 paired antibodies and 805 nanobodies.

\noindent \textbf{Downstream antibody functional dataset}
Assessing antibody function is of paramount importance in diagnostic and therapeutic applications. The generation of downstream functional datasets facilitates the evaluation of different methodologies' efficacy in comprehending immune responses, encompassing aspects such as binding specificity, paratope identification, redundancy, and the affinity of mutations.

 \textbf{Antigen binding} The CDR3 region of the clinically approved antibody trastuzumab sequence has been replaced with diverse variants. The dataset, obtained from \cite{Mason2021}, consists of 21,612 sequences, including 11,277 binders and 10,335 non-binders. These sequences are partitioned into 70\% for training, 15\% for validation, and 15\% for testing. Antibody sequences were truncated at the CDR3 region.

 \textbf{Paratope prediction} Human antibody sequence and structural data, annotated with antigen binding sites, were obtained from the SAbDab database \cite{SAbDab} (version 26 Aug 2021). Following the protocol established in \cite{antiberta}, only antigen-antibody complexes with resolutions below $\SI{3}{\angstrom}$ and binding to proteins or peptides were considered. After excluding single-chain structures and those missing more than two residues in the CDR regions, heavy and light chains were clustered using CD-HIT \cite{cdhit} at a $99\%$ identity threshold. The resulting 900 non-redundant sequences were divided into training, validation, and testing sets, with respective splits of 720, 90, and 90.

 \textbf{Redundancy prediction}  
The redundancy prediction dataset was compiled by filtering the OAS \cite{newOAS} dataset for sequences meeting these criteria: Chain = heavy, Isotype = IGHG, Species = human, Vaccine = None, Disease = SARS-CoV-2, B-Source = PBMC, Unique Sequences $\geq$ 10,000 and redundancy $\geq$ 3. For each subject, sequences within the 99th percentile were labeled as high redundancy, while those below the 10th percentile were categorized as low redundancy. To prevent overlap across subjects, we randomly distributed the resultant 42 subjects into three categories: 37 for training, 2 for validation and 3 for testing. We maintained all high redundancy sequences and randomly selected an equivalent number of low redundancy sequences from the same subject. Consequently, the final dataset composition comprised 27,764 sequences for training, 1,404 for validation and 2,550 for testing.

 \textbf{Affinity prediction}
The binding affinity training dataset for antibodies were obtained from the SKEMPI V2.0 \cite{SKEMPI} dataset, encompassing a comprehensive set of experimentally determined binding free energy changes for protein-protein complexes upon mutation. The testing set is composed of subsets from S1131 \cite{S1131} and M1707 \cite{M1707} datasets. Specifically, the S1131 dataset is a subset of SKEMPI, predominantly examining single-point mutations within antibody-antigen interactions. The M1707 dataset which is also a SKEMPI derivative probes binding free energy changes due to multi-point mutations in both wild-type and mutant SARS-CoV-2 proteins. Upon antibody selection, the training set includes 1388 entries extracted from the SKEMPI V2.0 dataset. The resultant testing set curated from S1131 and M1707 comprises 213 entries.

\noindent \textbf{Antibody structure prediction benchmark dataset}
\label{new_igfold_benchmark}
In order to conduct a rigorous comparison of the BALMFold's efficacy vis-à-vis alternate approaches, we utilized the latest IgFold benchmark \cite{igfold}. This benchmark dataset encompasses 197 paired and 71 nano antibody structures derived from the SAbDab database. The specified structures were made available between the temporal window of July 1, 2021, and September 1, 2022. Consistent with IgFold \cite{igfold}, structures exhibiting an excess of $\SI{3}{\angstrom}$ and those with a CDR H3 loop length surpassing 21 residues were systematically excluded.

\subsection{Training details}

\noindent \textbf{Pre-training objective}
The antibody vocabulary comprises 33 tokens, following the ESM-1b configuration \cite{esm1b}. The parameters of BALM are initialized using the protein language model ESM-2 checkpoint with 150M parameters, as released by Huggingface. Our antibody-specific language model employs the masked language modeling (MLM) objective \cite{bert} to predict masked amino acids. The MLM loss is defined as:

\begin{equation}
      \mathcal{L}_{\mathrm{MLM}}=\mathbb{E}_{s \sim S} \mathbb{E}_M \sum_{i \in M}-\log p\left(s_i \mid s_{/ M}\right).  
\end{equation}

The pre-training objective minimizes the negative log-likelihood of the actual residue, given the masked sequence. We generally adopted the BERT setting \cite{bert} to process selected tokens. Based on a predefined probability distribution of amino acids, 80\% of masked tokens were replaced by \textless mask{\textgreater}, 10\% by random amino acids, and the remaining 10\% of masked amino acids were left unchanged.

Pre-trained language models typically mask a certain percentage of tokens. The MLM performance depends on the proportion of masked tokens during the pre-training process. An appropriate mask ratio is crucial for ensuring the MLM effectively learns meaningful contextual representation from the corpus. Amino acid variability in antibody sequences differs between framework regions and CDRs, as illustrated in Fig. \ref{fig1}c. Predicting a conserved amino acid that has been masked is more straightforward than predicting a non-conserved one. Consequently, a constant masking probability distribution might hinder the extraction of CDR mutation knowledge from the antibody sequence. If masked positions are uniformly sampled, the relatively low loss on conserved locations would reduce the overall loss during the pre-training process. However, lower MLM loss in conserved regions does not necessarily indicate that the model captures essential antibody properties. Thus, masking different positions with varying probabilities is required.

Positions with a greater variety of categories in Fig. \ref{fig1}c should have higher masking probabilities than those with fewer categories. To align the mask probability distribution with the training corpus, we computed the entropy of the amino acid distribution at each position and rescaled the statistics to maintain a 15\% ensemble mask ratio. The entropy is calculated as:
\begin{equation}
{\rm Entropy} = -\sum_{i}p_{i}\log_{2}p_{i}.
\end{equation}

The entropy mask strategy enables the model to focus on learning patterns in both conserved and variable regions, enhancing its ability to capture the diverse properties of antibody sequences. In highly conserved sites with low computed entropy, the mask ratio may be extremely low, approaching zero. Considering the vital biological evolutionary information within these conserved regions, we adjusted the mask probability for positions with a mask probability below 10\% to 10\%. This adjustment ensures that the model can learn the broad biological properties of antibodies through amino acid restoration. We also set the masked probability to 20\% for insertions in CDR 3 not included in Fig. \ref{fig1}c. Based on this biologically meaningful regulation, the ensemble mask rate ranges from 17\% to 20\%, varying according to the length of sequences (complete masking probability for each position is presented in Supplementary Fig. \ref{mask_prob}).

\noindent \textbf{Hyperparameters for BALM pre-training}
 The AdamW optimizer \cite{adamw} was employed, featuring $\beta_1=0.9$ and $\beta_2 = 0.999$, a weight decay of 0.01, and the Noam learning rate scheduler \cite{transformer} with 0.17 warmup ratio. BALM underwent six epochs of training (12 days) on eight NVIDIA A100 GPUs, with a batch size of 256 per GPU. Owing to the well-annotated sequential data in the OAS dataset, biologically motivated position identifiers were readily obtainable. For unannotated single sequences, ANARCI \cite{anarci} with the IMGT numbering scheme was utilized for processing.

\noindent \textbf{BALMFold training details} 
BALMFold was trained with over 130K iterations utilizing eight A100 GPUs and a batch size of 32. The training procedure engaged an Adam optimizer with a learning rate of $3\times10^{-4}$ including weight decay and a warm-up ratio of 10\%. Initial 100K training iterations involved all single-chain antibodies from the dataset, incorporating discrete heavy and light chains extracted from paired antibodies. Successive 30K training iterations concentrated on paired antibodies to facilitate prediction of intricate antibody structures that include inter-chain orientations. 

\noindent \textbf{Fine-tune on downstream tasks}
The pre-trained language model  was independently fine-tuned for each unique task. Every model was subject to identical fine-tuning processes, with the exception of their learning rates. Optimal learning rates were determined using the validation set, where an exploration of fivefold learning rate values ranging from $1\times10^{-6}$ to $1\times10^{-3}$ was conducted, complemented by a $10\%$ warmup schedule. To assess the generalizability and stability of the models, subsequent evaluations on the test set were performed using three distinct random seeds. The resultant average values and standard deviations are reported.

\subsection{Evolutionary velocity of antibody language model}
The vector field of immune system responses to specific antigen is considered to be evolutionary velocity of antibody. Based on pseudo log-likelihoods \cite{persudolikelihood}, evo-velocity \cite{evolocity} score between sequence $s^{a}$ to sequence $s^{b}$ is computed as:
\begin{equation}
        v_{a b} {=} \frac{1}{\vert\mathcal{D}\vert} \sum_{i \in \mathcal{D}}\left[\log p\left(s_i^{b} \mid \mathbf{z}_i^{a}\right)-\log p\left(s_i^{a} \mid \mathbf{z}_i^{b}\right)\right].
\end{equation}
Here, $\mathcal{D} = \{i:s_{i}^{a}\neq s_{i}^{b}\}$ represents the position set of different residues between two sequences after pairwise sequence alignment, and $\mathbf{z}_i$ indicates the representation of sequence excluding residue on position $i$.

\subsection{Calculation of performance metrics} 
Subtle variations in residues at each position within the CDRs of the heavy chain may lead to distinct antigen-binding properties. To understand specific antibody functions and explore the capability of the aligned antibody language model in identifying novel therapeutics, we executed four downstream tasks—antigen binding, paratope prediction, redundancy classification, and affinity regression. For the assessment of classification performance on three downstream tasks, we calculated true positive (TP), true negative (TN), false positive (FP), and false negative (FN) values to determine performance metrics such as precision (P), recall (R), F1 score, Matthews' correlation coefficient (MCC), area under the receiver operating characteristic curve (AUC), and average precision scores with recall (APR). The Pearson's $r$ ($r$) was employed to evaluate the regression task. A higher Pearson's $r$-value signifies a strong correlation between predicted and actual values.  Correspondingly, the computational formulas are summarized as follows:
$$
   {\rm P} = \frac{{\rm TP}}{{\rm TP}+{\rm FP}},~~
   {\rm R} = \frac{{\rm TP}}{{\rm TP}+{\rm FN}},
~~
    {\rm F}1 = \frac{2{\rm P}\times {\rm R}}{{\rm P}+{\rm R}},
$$
$$
   {\rm MCC} = \frac{\rm TP\times TN-FP\times FN}{\sqrt{({\rm TP+FP})({\rm FN+TP})({\rm FN+TN})({\rm FP+TN})}},
$$
$$
    {\rm APR} = \sum_{n}P_{n}(R_{n}-R_{n-1}),
$$
and
$$
    r = \frac{\sum\limits_{i=1}^n (x_i - \bar{x})(y_i - \bar{y})}{\sqrt{\sum\limits_{i=1}^n (x_i - \bar{x})^2} \sqrt{\sum\limits_{i=1}^n (y_i - \bar{y})^2}}.
$$

\section{Data Availability}
The datasets utilized in this project are publicly available. The antibody sequences used for the pre-training of the language model were sourced from the OAS database \url{https://opig.stats.ox.ac.uk/webapps/oas/}. The paratope prediction dataset was download from \url{https://github.com/alchemab/antiberta} and the antigen binding prediction dataset was obtained from \url{https://zenodo.org/record/7340488}. The affinity prediction task was conducted using SKEMPI 2.0 \url{https://life.bsc.es/pid/skempi2}. Antibody structures for the training process were retrieved from the SAbDab database \url{https://opig.stats.ox.ac.uk/webapps/newsabdab/sabdab/}. The target protein data bank (PDB) ids for the antibody structure prediction benchmark were obtained from the Zenodo database \url{https://doi.org/10.5281/zenodo.7677723}, with the corresponding experimentally determined structures fetched from the SAbDab dataset.

\section{Code Availability}
The BALMFold structure prediction server can be reached at \url{https://beamlab-sh.com/models/BALMFold}. The inference code for BALM is made available at \url{https://github.com/BEAM-Labs/BALM}. The pre-trained weights for BALM can also be obtained from the aforementioned GitHub repository. The pre-training code for the model is derived from HuggingFace \url{https://huggingface.co/facebook/esm2\_t30\_150M\_UR50D}. For IMGT numbering, the ANARCI tool is available at \url{https://opig.stats.ox.ac.uk/webapps/newsabdab/sabpred/anarci/}.

\bibliographystyle{recomb}  
\bibliography{main}
\newpage
\section*{Supplementary Information} 
\setcounter{figure}{0}
\setcounter{table}{0}
\renewcommand{\figurename}{Supplementary Fig.}
\renewcommand{\tablename}{Supplementary Table}

\begin{figure}[H]
	\centering
    \includegraphics[scale=0.22]{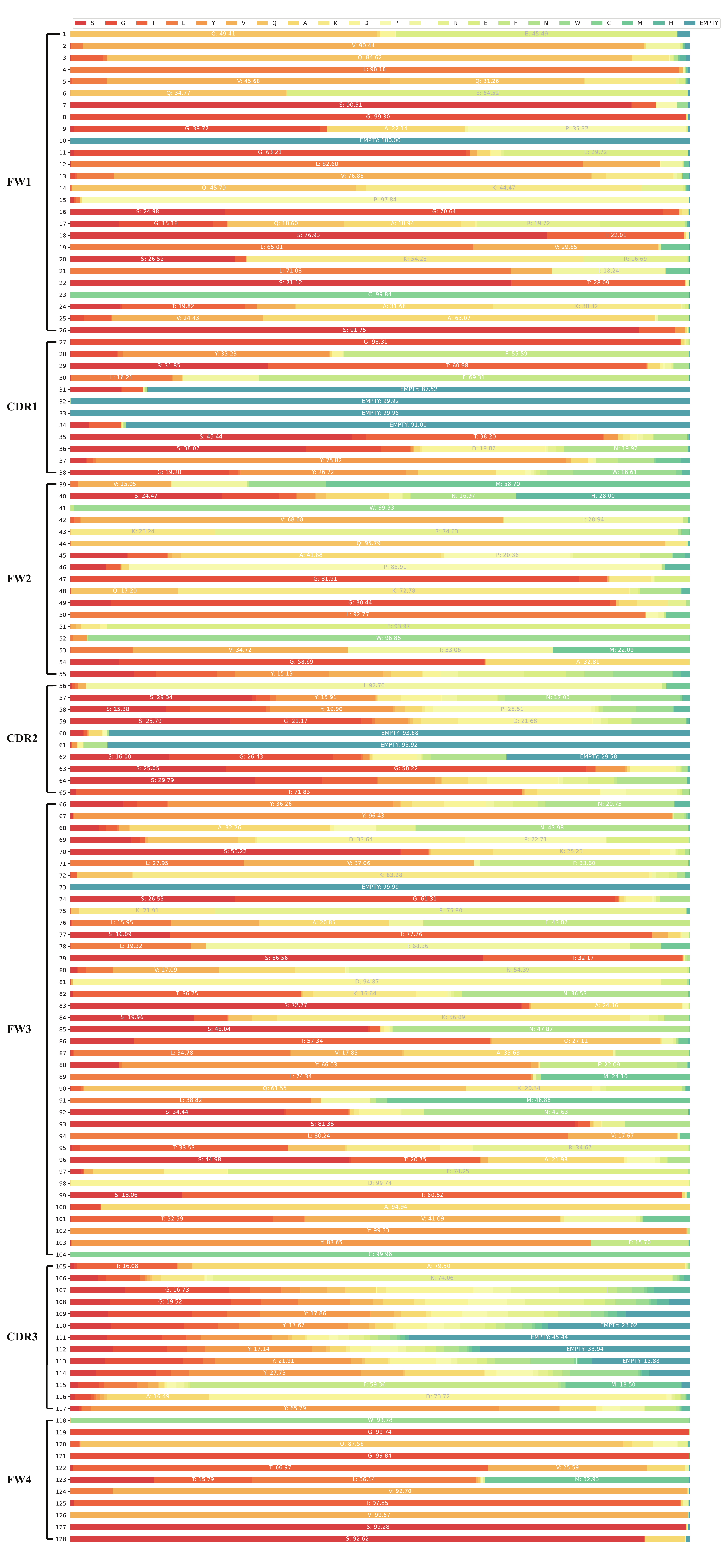}
	\caption{Amino acids distribution of the heavy chains of all paired antibody sequences from the OAS database.}
	\label{aa_freq}
\end{figure}

\newpage

\begin{figure}[H]
	\centering
    \includegraphics[width=\linewidth]{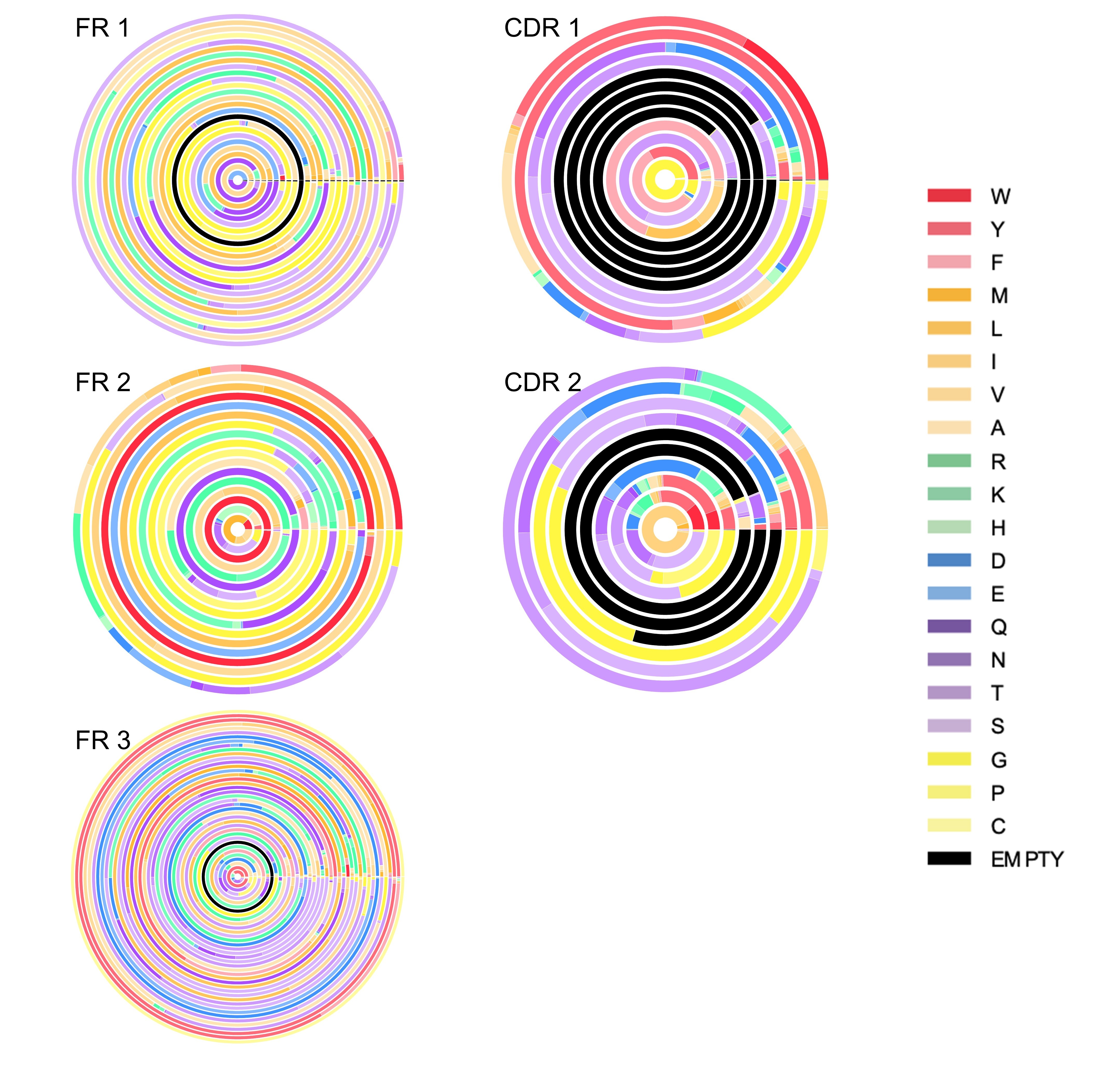}
	\caption{Depiction of the distribution of amino acids across different positions within heavy chains from the OAS paired dataset.}
	\label{all_aa_distribution}
\end{figure}

\newpage

\begin{figure}[H]
	\centering
    \includegraphics[scale=0.45]{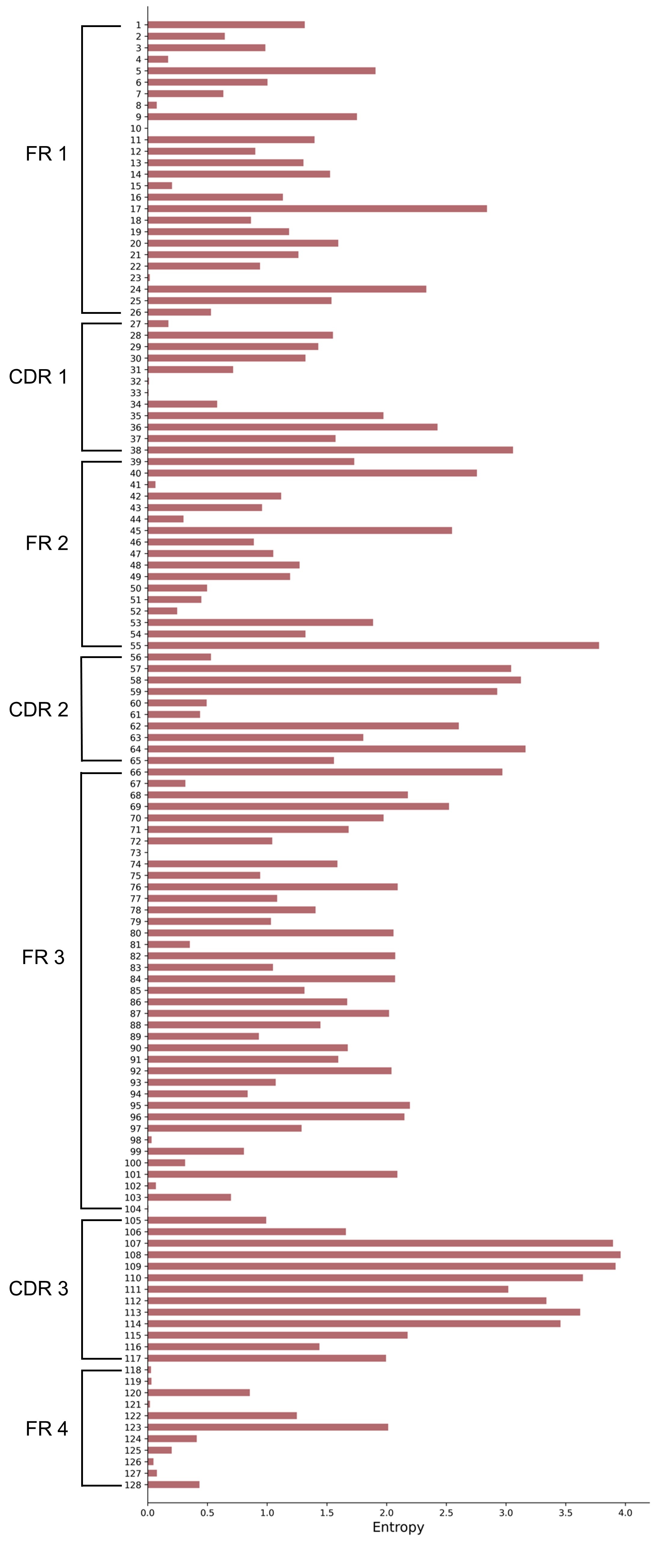}
	\caption{Illustration of the distribution of amino acid entropy across varied positions within antibody sequences.}
	\label{all_entropy}
\end{figure}

\newpage

\begin{figure}[H]
	\centering
    \includegraphics[scale=0.45]{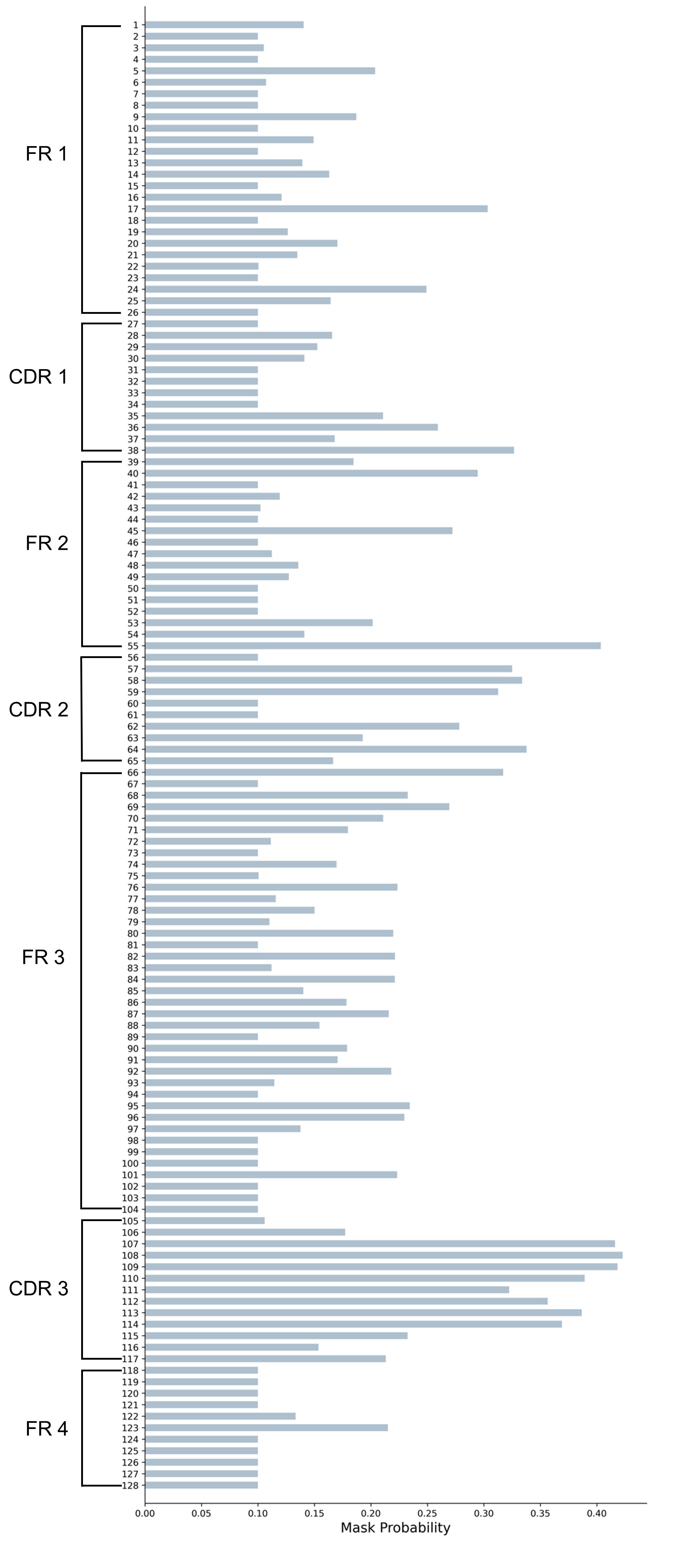}
	\caption{Illustration of the probability distribution of the masking strategy across various positions within antibody sequences during the pre-training stage.}
	\label{mask_prob}
\end{figure}

\newpage

\begin{figure}[H]
	\centering
    \includegraphics[width=\linewidth]{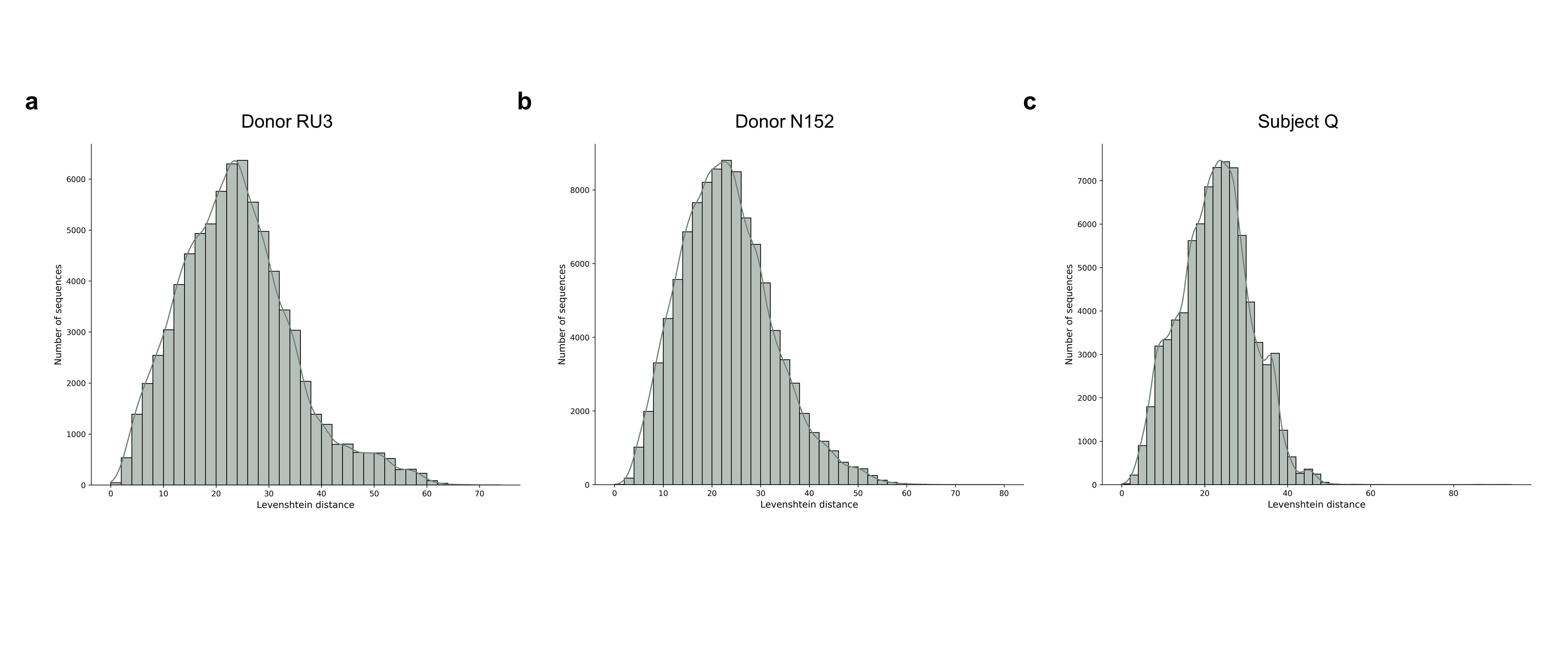}
	\caption{
     \textbf{Distribution of Levenshtein distance between germline sequences and antibody sequences in three donors.}
 }
	\label{LD}
\end{figure}

\newpage

\begin{table}[h]
\caption{Comparison of language model architectures. AntiBERTy with 26M and ESM-2 with 150M parameters were compared.}
\centering
\setlength{\tabcolsep}{0.5mm}
{
\begin{tabular}{lcccccc}
\hline &Num Layer & Hidden state & Attention head & Intermediate dim & Num params & Database\\
\hline 
AntiBERTy & $8$ & $512$ & $8$ & $2048$ & $26M$ & OAS (558M)\\
AbLang-H  & $12$ & $768$ & $12$ & $3072$  & $86M$ & OAS (14M)\\
EATLM & $12$ & $768$ & $12$ & $3072$  & $86M$ & OAS  (20M)\\
AntiBERTa  & $12$ & $768$ & $12$ & $3072$  & $86M$& OAS  (58M)\\
ESM-2  & $30$ & $640$ & $20$ & $2560$ & $150M$ & UniRef50 \\
ESM-1b  & 33 & 1280 & 20 & 5120 & $650M$ & UniRef50 \\ 
BALM  & $30$ & $640$ & $20$ & $2560$ & $150M$ & OAS (336M) \\
\hline
\end{tabular}}
\label{model arch}
\end{table}

\newpage

\begin{table}[h]
    \centering
    \caption{Affinity maturation prediction with Pearson's $r$ among baselines. Language models with pre-trained and without pre-trained are compared.}
    \setlength{\tabcolsep}{1mm}{
    \begin{tabular}{lc}
    
    \hline 
           Model & Pearson's r\\
    \hline 
           BALM & $87.4\pm0.6$\\
           BALM w/o pre-trained & $72.2\pm7.7$\\
           ESM-2 & $85.9\pm0.6$\\
           ESM-2 w/o pre-trained & $66.6\pm12.1$\\
           AntiBERTy & $60.5\pm2.5$\\
          AntiBERTy w/o pre-trained & $48.0\pm0.9$\\
    \hline 
    \end{tabular}}
    \label{affinity}
\end{table}

\newpage

\begin{table}[h]
\centering
\caption{Comparison of structure prediction performance of methods with average RMSD scores on paired antibody benchmark.}
\label{tab:performance-new-IgFold-benchmark-pair}
\setlength{\tabcolsep}{1mm}{
\begin{tabular}{l|c|cccc|cccc}
\hline
Method &
\multicolumn{1}{{|c|}}{OCD} &
\multicolumn{1}{c}{CDR H1} &
\multicolumn{1}{c}{CDR H2} &
\multicolumn{1}{c}{CDR H3} &
\multicolumn{1}{c|}{FR H} &
\multicolumn{1}{c}{CDR L1} &
\multicolumn{1}{c}{CDR L2} &
\multicolumn{1}{c}{CDR L3} &
\multicolumn{1}{c}{FR L} \\ \hline
BALMFold &
\textbf{3.31} &
\textbf{0.77} &
\textbf{0.60} &
\textbf{3.05} &
\textbf{0.38} &
\textbf{0.59} &
\textbf{0.37} &
\textbf{0.94} &
\textbf{0.35} \\

AlphaFold2-M & 4.18  & 0.95 & 0.74 & 3.56 & 0.69 & 0.84 & 0.51 & 1.59 & 0.66 \\
IgFold & 3.84 & 0.91 & 0.82 & 3.42 & 0.50 & 0.79 & 0.51 & 1.22 & 0.47
 \\
ESMFold & - &0.94 & 0.94 & 4.41 & 0.50 & 1.01 & 0.53 & 1.59 & 0.45
 \\ 
OmegaFold &- & 0.89 & 0.72 & 3.55 & 0.47 & 0.73 & 0.41 & 1.17 & 0.41
\\
 \hline
\end{tabular}}
\label{table: pair benchamrk}
\end{table}

\newpage

\begin{table}[h]
\centering
\caption{Comparison of structure prediction performance of methods with average RMSD scores on nanobody benchmark.}
\label{tab:performance-new-IgFold-benchmark-nano}
\setlength{\tabcolsep}{2mm}{
\begin{tabular}{lcccc}
\hline
Method &
\multicolumn{1}{l}{CDR 1} &
\multicolumn{1}{l}{CDR 2} &
\multicolumn{1}{l}{CDR 3} &
\multicolumn{1}{l}{FR} \\ \hline
BALMFold &
\textbf{1.50} &
\textbf{0.83} &
\textbf{3.69} &
\textbf{0.53}
 \\
AlphaFold2 & 1.61 & 0.88 & 4.00 & 0.57
 \\
IgFold & 1.80 & 1.10 & 4.31 & 0.63
 \\
ESMFold & 1.73 & 0.87 & 4.78 & 0.58
 \\ 
 OmegaFold & 1.79 & 0.88 & 4.05 & 0.57
 \\ 
 \hline
\end{tabular}}
\label{table: nano benchamrk}
\end{table}

\newpage

\begin{figure}[H]
	\centering
    \includegraphics[width=\linewidth]{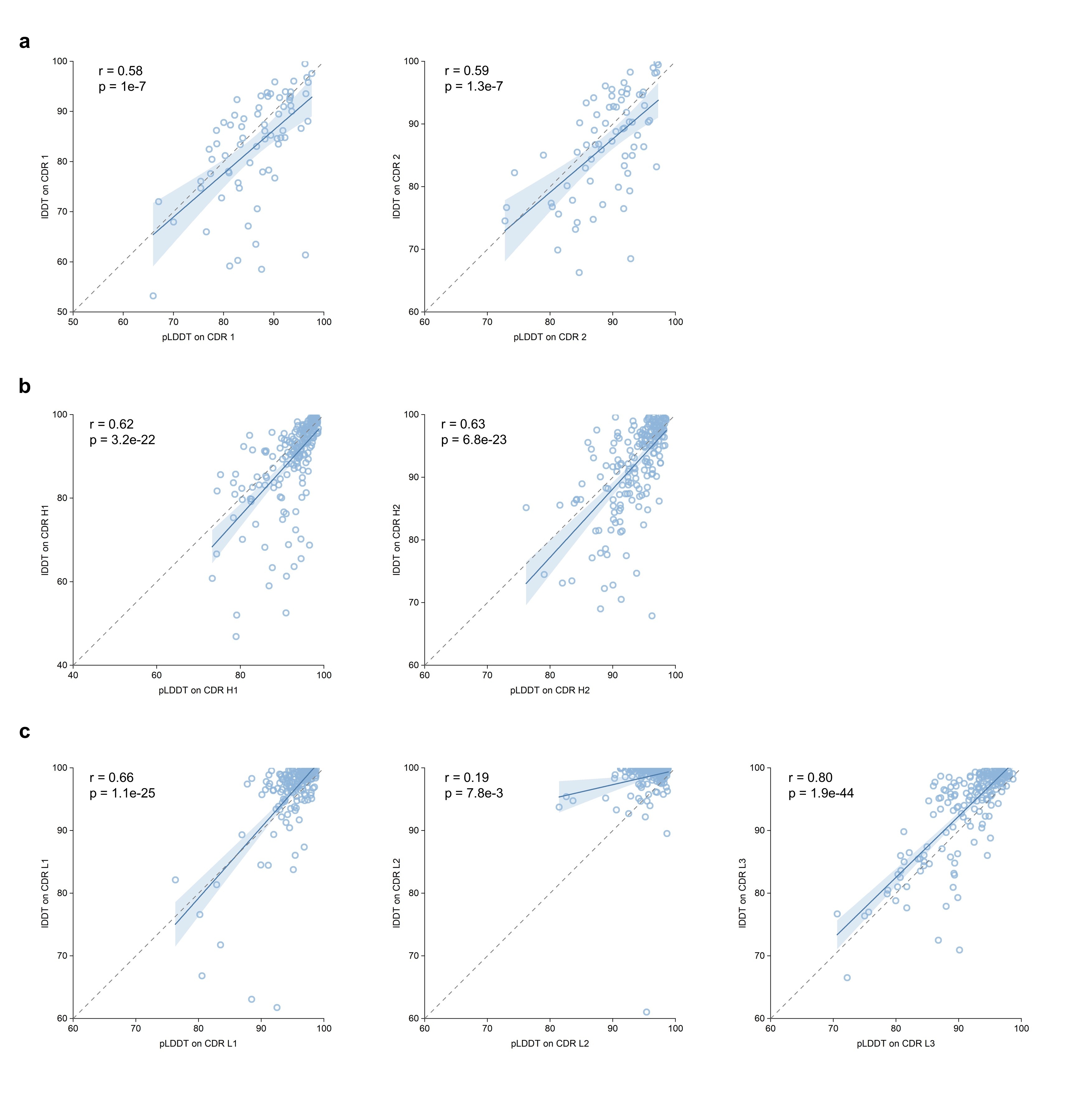}
	\caption{
    \textbf{Correlation between confidence score pLDDT and true lDDT on benchmark.}
    \textbf{a}, Correlation on CDR 1 and CDR 2 of nanobodies. Linear regression fits each samples, illustrated with Pearson's $r$ score and P-value on the top of left. The shaded area represents the 95\% confidence interval.
    \textbf{b}, Correlation on heavy chain of paired antibodies.
    \textbf{c}, Correlation on light chain of paired antibodies.
    }
    \label{figure: plddt with lddt}
\end{figure}

\newpage

\begin{figure}[H]
	\centering
    \includegraphics[width=\linewidth]{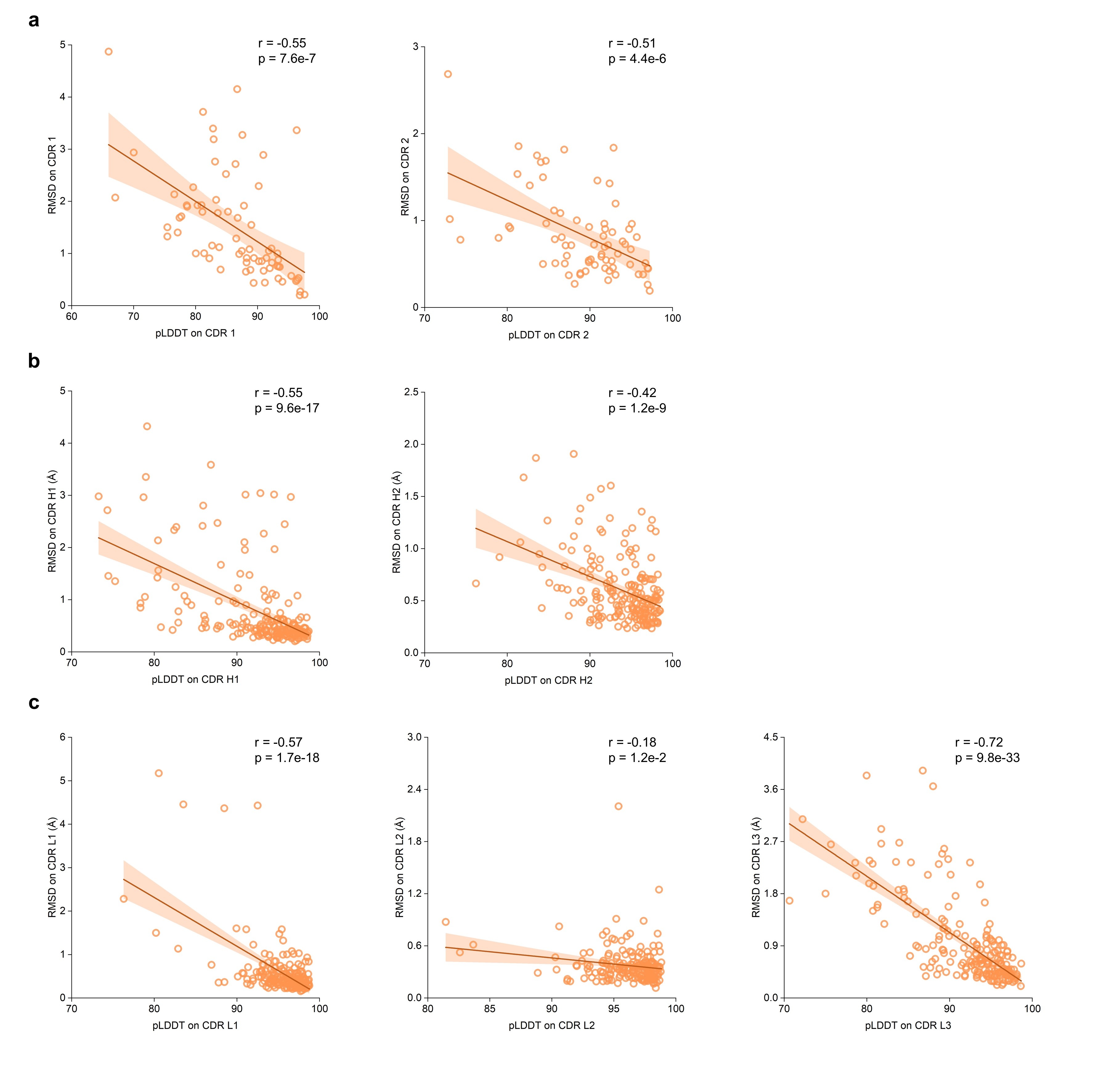}
	\caption{
    \textbf{Correlation between confidence score pLDDT and true RMSD value on benchmark.}
    \textbf{a}, Correlation on CDR 1 and CDR 2 of nanobodies. Each dots represent a sample. Linear regression fits each samples, illustrated with Pearson's $r$ score and P-value on the top of right. The shaded area represents the 95\% confidence interval.
    \textbf{b}, Correlation on heavy chain of paired antibodies.
    \textbf{c}, Correlation on light chain of paired antibodies.
    }
    \label{figure: plddt with rmsd}
\end{figure}

\newpage

\begin{figure}[H]
	\centering
    \includegraphics[width=\linewidth]{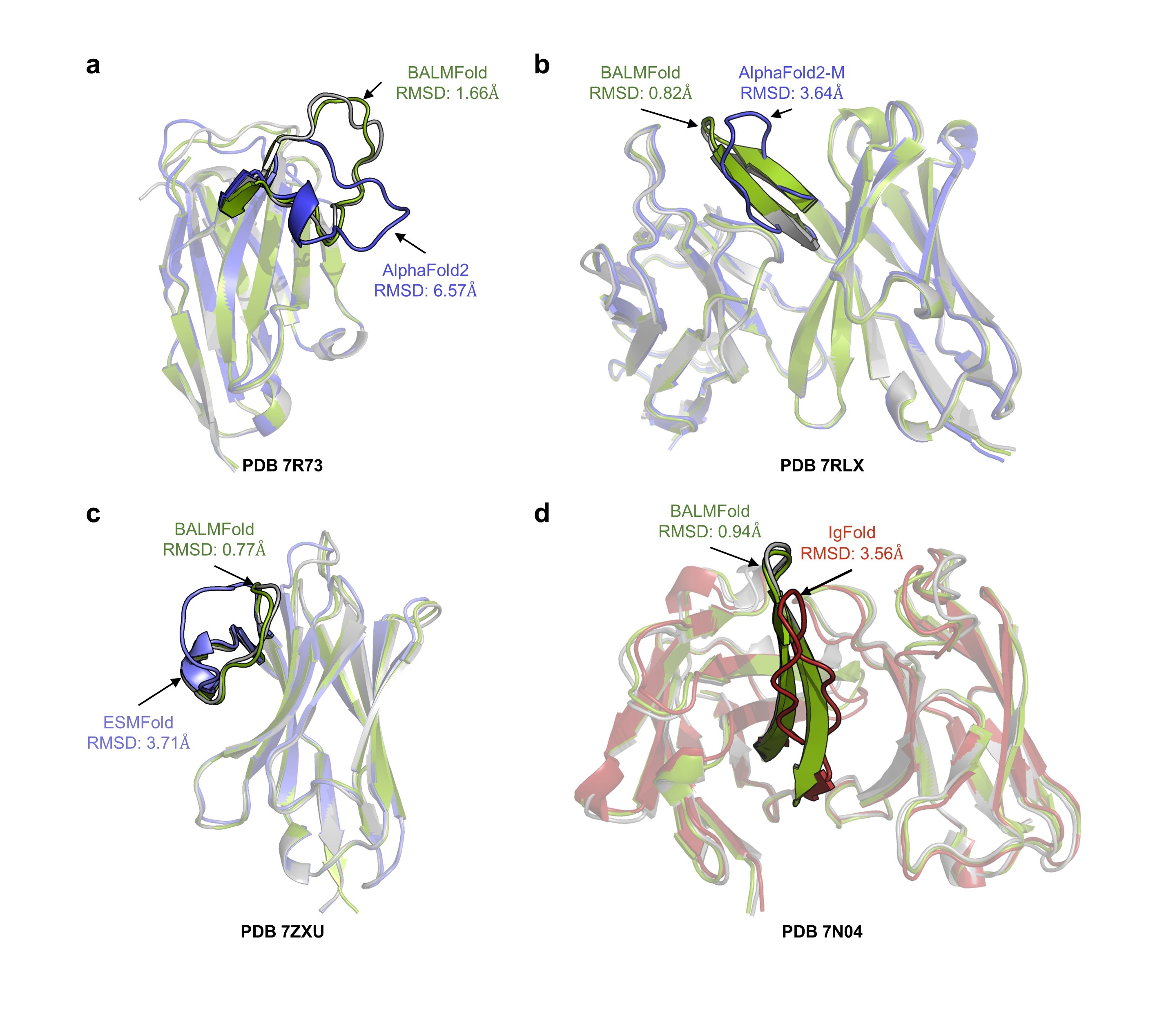}
	\caption{
    \textbf{Case studies in the structure prediction benchmark.} The native experimental structures are colored in grey.
    \textbf{a}, Illustration of structure predictions by BALMFold (green) and AlphaFold2 (blue) for target 7R73 ($\mathcal{L}_\text{CDR H3}=18$ residues). 
    \textbf{b}, Illustration of structure predictions by BALMFold (green) and AlphaFold2-M (blue) for target 7RLX ($\mathcal{L}_\text{CDR H3}=12$ residues).
    \textbf{c}, Illustration of structure predictions by BALMFold (green) and ESMFold (light blue) for target 7ZXU ($\mathcal{L}_\text{CDR H3}=12$ residues).
    \textbf{d}, Illustration of structure predictions by BALMFold (green) and IgFold (red) for target 7N04 ($\mathcal{L}_\text{CDR H3}=18$ residues).
    }
        \label{figure: more case study}

\end{figure}

\newpage

\begin{figure}[H]
	\centering
    \includegraphics[scale=0.75]{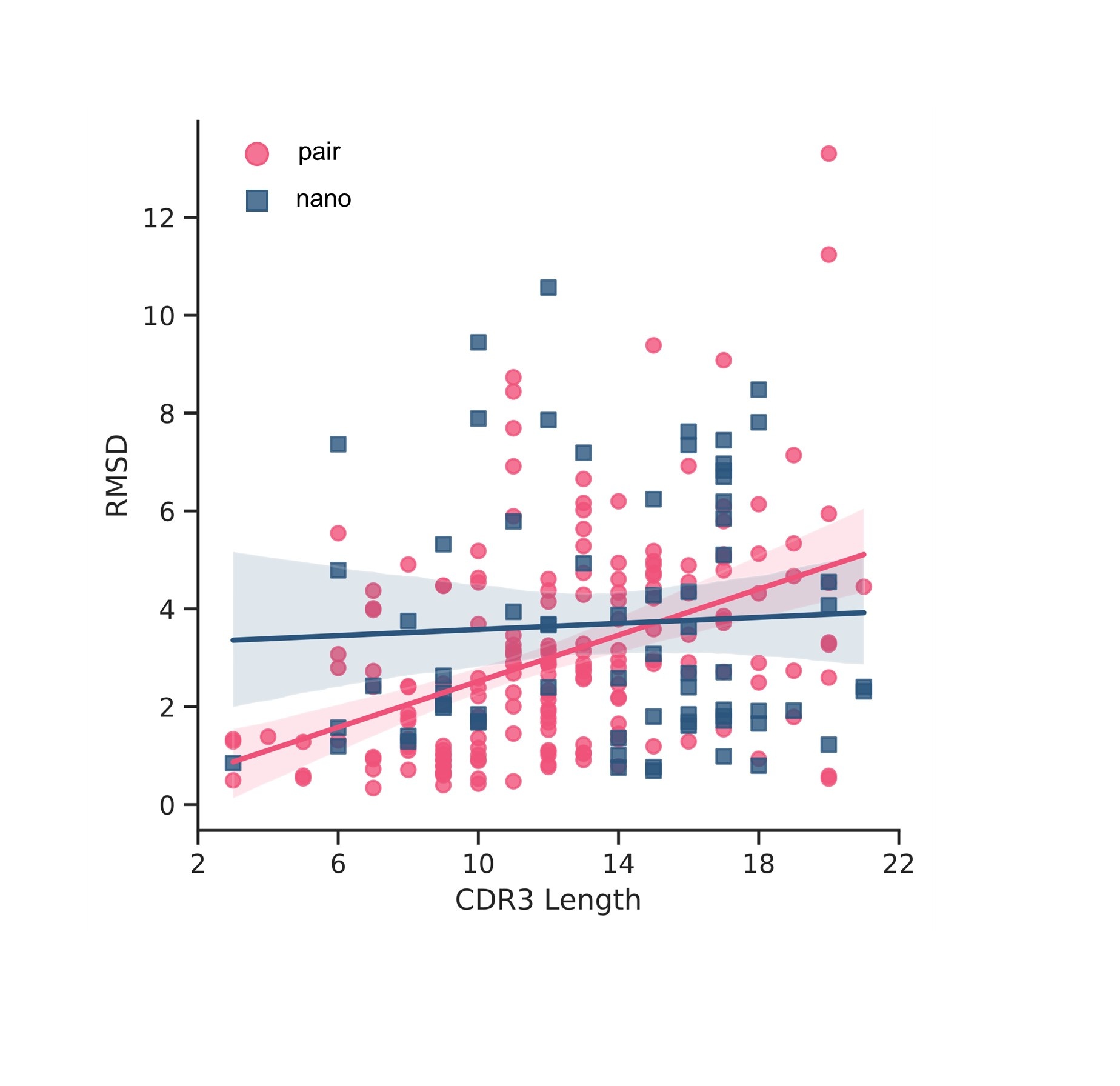}
	\caption{
        \textbf{Regression analysis between RMSD and the length of the CDR H3 loop for both paired and nano antibodies.}
         Each point corresponds to a single antibody prediction structure of BALMFold.
         }
    \label{figure: rmsd vs cdr3 length}
\end{figure}

\newpage

\noindent \textbf{Supplementary benchmark with 143 paired antibodies and 57 nanobodies.}
We compiled a selection of experimentally determined antibody structures, made available between July 1st, 2021 and August 1st, 2022 from SAbDab \citep{SAbDab}, employing an analogous processing approach to that used for the training dataset. After the exclusion of samples included in the earlier IgFold benchmark \citep{igfold} (67 paired and 21 single-chain antibodies), the final benchmark encompassed 143 paired antibodies and 57 nanobodies. This supplementary benchmark contains 93 paired antibodies and 35 nanobodies that are congruent with the benchmark detailed in the main text \ref{new_igfold_benchmark}.

\newpage

\begin{table}[h!]
\centering
\caption{Comparison of structure prediction performance of methods with average atomic RMSD scores on additional benchmark with 57 nanobodies.}
\label{tab:performance-balmfold-benchmark-nano}
\setlength{\tabcolsep}{2mm}{
\begin{tabular}{lcccc}
\hline
Method &
\multicolumn{1}{l}{CDR 1} &
\multicolumn{1}{l}{CDR 2} &
\multicolumn{1}{l}{CDR 3} &
\multicolumn{1}{l}{FR} \\ \hline
BALMFold &
\textbf{1.53} &
\textbf{0.83} &
\textbf{3.09} &
0.61
 \\
AlphaFold2 & 1.74 & 0.97 & 4.12 & 0.58
 \\
IgFold & 1.78 & 1.10 & 4.28 & 0.71
 \\
ESMFold & 1.71 & 0.94 & 3.88 & 0.64
 \\ 
 OmegaFold & 1.70 & 0.91 & 3.25 & \textbf{0.56}
 \\ 
 \hline
\end{tabular}}
\end{table}

\newpage

\begin{table}[h!]
\centering
\caption{Comparison of structure prediction performance of methods with average atomic RMSD scores on additional benchmark with 143 paired antibodies.}
\label{tab:performance-BALMFold-benchmark-pair}
\setlength{\tabcolsep}{1mm}{
\begin{tabular}{l|c|cccc|cccc}
\hline
Method &
\multicolumn{1}{{c|}}{OCD} &
\multicolumn{1}{c}{CDR H1} &
\multicolumn{1}{c}{CDR H2} &
\multicolumn{1}{c}{CDR H3} &
\multicolumn{1}{c|}{FR H} &
\multicolumn{1}{c}{CDR L1} &
\multicolumn{1}{c}{CDR L2} &
\multicolumn{1}{c}{CDR L3} &
\multicolumn{1}{c}{FR L} \\ \hline
BALMFold &
\textbf{3.71} &
\textbf{0.82} &
\textbf{0.71} &
\textbf{3.32} &
\textbf{0.41} &
\textbf{0.63} &
\textbf{0.49} &
\textbf{1.05} &
\textbf{0.40} \\
IgFold & 4.91 & 1.00 & 0.91 & 3.79 & 0.53 & 0.93 & 0.68 & 1.38 & 0.55 \\
AlphaFold2 & - & 0.90 & 0.84 & 4.07 & 0.50 & 0.84 & 0.62 & 1.30 & 0.50 \\
OmegaFold & -& 0.91 & 0.79 & 3.92 & 0.48 & 0.82 & 0.54 & 1.25 & 0.46 \\
ESMFold & -& 0.98 & 0.97 & 4.20 & 0.57 & 1.01 & 0.63 & 1.52 & 0.49 \\ \hline
\end{tabular}}
\end{table}

\newpage

\begin{table}[h!]
\centering
\caption{Comparison of structure prediction performance of methods with average atomic RMSD scores on earlier IgFold \citep{igfold} benchmark with 67 paired antibodies.}
\label{tab:performance-IgFold-benchmark-pair}
\setlength{\tabcolsep}{1mm}{
\begin{tabular}{l|c|cccc|cccc}
\hline
Method &
\multicolumn{1}{{c|}}{OCD} &
\multicolumn{1}{c}{CDR H1} &
\multicolumn{1}{c}{CDR H2} &
\multicolumn{1}{c}{CDR H3} &
\multicolumn{1}{c|}{FR H} &
\multicolumn{1}{c}{CDR L1} &
\multicolumn{1}{c}{CDR L2} &
\multicolumn{1}{c}{CDR L3} &
\multicolumn{1}{c}{FR L} \\ \hline
BALMFold &
\textbf{3.39} &
\textbf{0.64} &
\textbf{0.58} &
\textbf{2.96} &
\textbf{0.38} &
\textbf{0.78} &
\textbf{0.39} &
\textbf{0.94} &
\textbf{0.34} \\

IgFold & 4.43 & 0.82 & 0.83 & 3.14 & 0.52 & 0.93 & 0.59 & 1.22 & 0.50 \\
AlphaFold2 & - & 0.73 & 0.76 & 3.62 & 0.47 & 0.99 & 0.49 & 1.35 & 0.45 \\
OmegaFold & - & 0.76 & 0.74 & 3.14 & 0.46 & 0.91 & 0.41 & 1.25 & 0.41 \\
ESMFold & - & 0.81 & 0.87 & 3.48 & 0.51 & 1.12 & 0.58 & 1.45 & 0.43 \\ \hline
\end{tabular}}
\end{table}

\end{document}